\documentclass[%
 reprint,
 amsmath,amssymb,
 aps,
citeautoscript]{revtex4-2}

\setcitestyle{super}

\newcommand*{\citen}[1]{%
  \begingroup
    \romannumeral-`\x % remove space at the beginning of \setcitestyle
    \setcitestyle{numbers}%
    \cite{#1}%
  \endgroup   
}

\usepackage{graphicx}% Include figure files
\usepackage{dcolumn}% Align table columns on decimal point
\usepackage{bm}% bold math

\begin{document}

\title{
Fast multi-source nanophotonic simulations using augmented partial factorization
}
\author{Ho-Chun Lin}
\author{Zeyu Wang}
\author{Chia Wei Hsu}
 \email{cwhsu@usc.edu}
\affiliation{%
Ming Hsieh Department of Electrical and Computer Engineering, University of Southern California, Los Angeles, California 90089, USA
}%

\begin{abstract}
Full-wave simulations are indispensable for nanophotonics and electromagnetics but are severely constrained on large systems, especially multi-channel ones such as disordered media, aperiodic metasurfaces, and densely packed photonic circuits where each input requires a large-scale simulation. 
Here we bypass the computationally demanding solution of Maxwell's equations and directly evaluate the full-wave multi-input response, with no approximation.
We augment the Maxwell operator with all input source profiles and output projection profiles, followed by a single partial factorization that directly yields the entire multi-input scattering matrix via the Schur complement.
This method is simple to implement and applies to any linear partial differential equation.
Its advantage grows with size, being 1,000 to 30,000,000 times faster than existing methods for systems with about ten million variables. % when volume discretized. %, while also using less memory.
We use it to realize the first full-wave simulations of entangled-photon backscattering from disorder and all-angle characterizations of high-numerical-aperture metalenses that are thousands of wavelengths wide.
This work reveals the significant efficiency gain when we rethink what to compute and enables the exploration of diverse multi-channel systems.
\end{abstract}

\maketitle

%\section{\label{sec:intro}Introduction}

\noindent
The interaction between light and nanostructured materials leads to rich properties. 
For small systems such as individual nano-/micro-structures and optical components, or for periodic systems such as photonic crystals and metamaterials that can be reduced to unit cells, one can readily solve Maxwell’s equations numerically to obtain predictions that agree quantitatively with experiments.
But the computation costs are typically too heavy for more complex systems such as disordered ones~\cite{Akkermans2007, carminati_schotland_2021} which not only are large but also couple many incoming channels to many outgoing ones, requiring numerous simulations.
The alternatives all have limitations: Born approximation does not describe multiple scattering, radiative transport and diagrammatic methods can only compute certain ensemble-averaged properties~\cite{Akkermans2007, carminati_schotland_2021}, and coupled-mode theory is restricted to systems with few isolated resonances~\cite{2004_Suh_JQE,2021_Zhou_ACSphoton_2}.
For metasurfaces~\cite{Yu2014_NM, Kamali2018_Nanophotonics}, the widely used locally periodic approximation~\cite{Pestourie2018_OE,Kamali2018_Nanophotonics}
is inaccurate whenever the cell-to-cell variation is large~\cite{Hsu2017_OE,2019_Phan_LSA,Lin2019_OE,Torfeh2020_ACSP, 2022_Skarda_CM} % 2018_Perez-Arancibia_OE, An2021_ArXiv
and cannot describe nonlocal responses~\cite{2020_Overvig_PRL,Spagele2021_NC} %2018_Kwon_PRL
and metasurfaces that are not based on unit cells~\cite{2020_Elsawy_LPR_review, Chung2020_OE, Lin2021_APL}.
Classical~\cite{2017_Shen_nphoton, 2020_Wetzstein_Nature_perspective} and quantum~\cite{2021_Pelucchi_NRP,2022_Madsen_Nature} photonic circuits build on individual components that couple very few channels at a time, limiting the number of inputs and outputs.
Examples beyond photonics also abound.
A wide range of studies across different disciplines are currently prohibited by computational limitations.

\begin{figure*}
\includegraphics[width=0.67\textwidth]{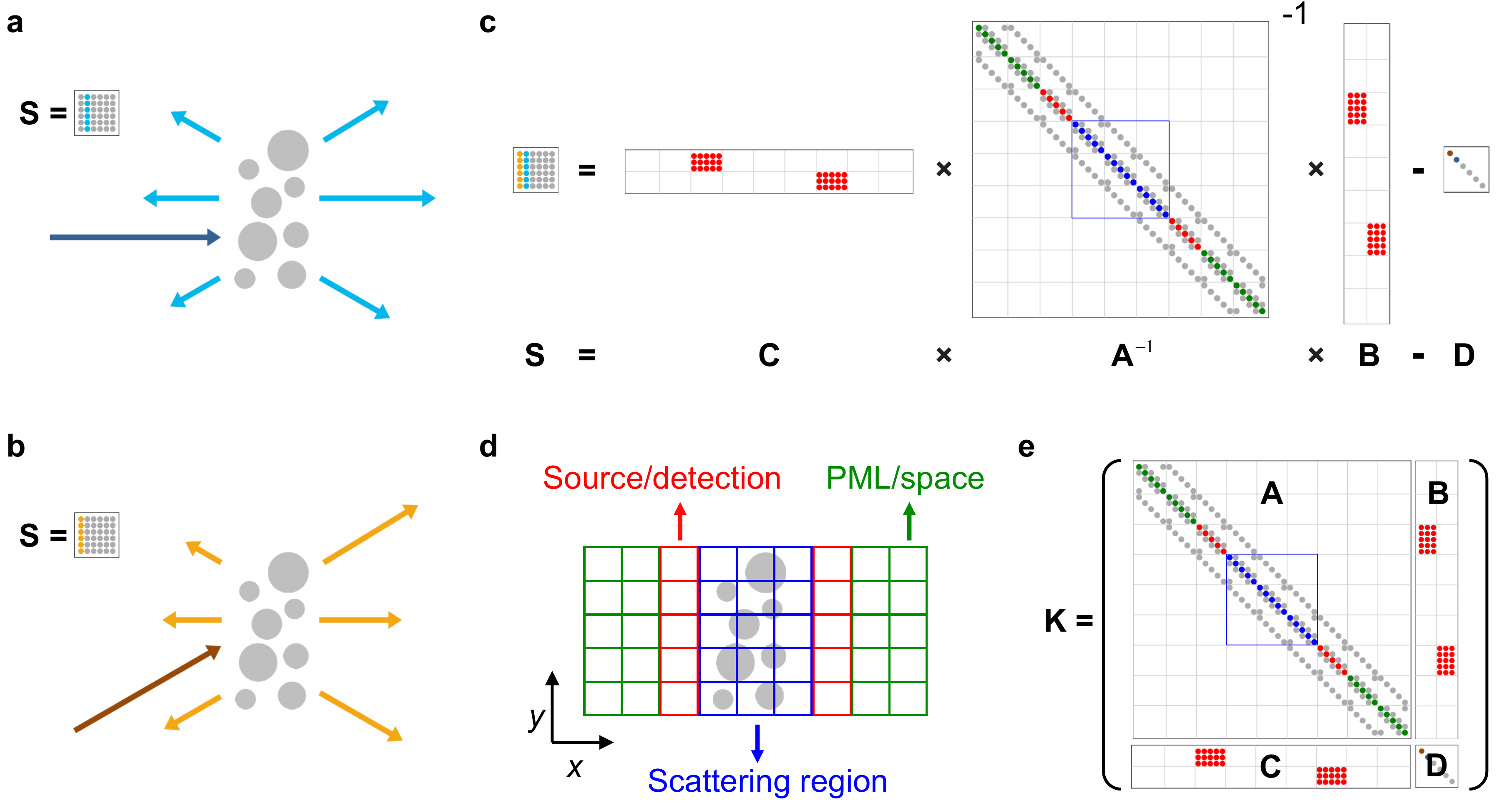}
\caption{\label{fig:schematic} {\bf Generalized scattering matrix and augmented partial factorization (APF).} 
{{\bf{a}}--{\bf{b}}}, Inputs from two different angles, corresponding to two columns of the generalized scattering matrix ${\bf S}$. {\bf{c}}, Illustration of Eq.~\eqref{eq:S}, which relates {\bf{S}} to the Maxwell operator ${\bf A}$, source profiles ${\bf B}$ that generate the incident waves, projection profiles ${\bf C}$ that extract the outputs of interest, and matrix ${\bf D}$ that subtracts the baseline. 
{\bf{d}}, The system used for the illustration in {\bf c}, with color coding for different parts of the system. 
{\bf{e}}, The augmented sparse matrix $\bf{K}$ of Eq.~\eqref{eq:K}, whose partial factorization gives the scattering matrix {\bf{S}}.}
\end{figure*}

Regardless of the complexity of a system, its linear response is described exactly by a generalized $M' \times M$ scattering matrix $\bf{S}$ that relates an arbitrary input vector $v$ to the resulting output vector $u$ via~\cite{2010_Popoff_PRL,Mosk2012_NP, Rotter2017_RMP,2019_Miller_AOP} % 2000_Carminati_PRA
\begin{equation}
{u_n} = \sum_{m=1}^M {{S_{nm}}{v_m}}. 
 \label{eq:S_eq}
\end{equation} 
The $M$ columns of $\bf{S}$ correspond to $M$ distinct inputs, illustrated in Fig.~\ref{fig:schematic}a--b.
The elements of the input vector $v$ can be the amplitudes of incoming channels in the far field or in waveguides, or point-dipole excitations in the near field, or any other input of interest; similarly, vector $u$ can contain any output of interest.

Computing such a multi-input response typically requires $M$ distinct solutions of Maxwell's equations with the same structure but with different source profiles.
Time-domain methods~\cite{2005_FDTD_book} are easy to parallelize but cannot leverage the multi-input property.
Frequency-domain methods allow strategies for handling many inputs.
After volume discretization through finite element~\cite{Jin2014} or finite difference~\cite{Shin2013, Rumpf2012_PIER}, Maxwell's equations in frequency domain becomes a system of linear equations ${\bf{A}}x_m=b_m$; sparse matrix ${\bf A}$ is the Maxwell differential operator, the right-hand-side column vector $b_m$ specifies the $m$-th input, and the solution is contained in column vector $x_m$.
When solving for $x_m={\bf{A}}^{-1}b_m$ with direct methods, the sparsity can be utilized via graph partitioning, and the resulting LU factors can be reused among different inputs~\cite{Davis2006, Duff2017}. But $M$ forward and backward substitutions are still needed, and the LU factors take up substantial memory.
Iterative methods compute $x_m = {\bf{A}}^{-1}b_m$ by minimizing the residual~\cite{Saad2003}, avoiding the LU factors. 
One can iterate multiple inputs together~\cite{Puzyrev2015_GJI} or construct preconditioners to be reused among different inputs~\cite{Dolean2015, 2016_Osnabrugge_JCP}, % Cai1999_SIAM, Saad2003, Dolean2009_SIAM, Smith2017
but the iterations still take $\mathcal{O}(M)$ time.

The boundary element method (BEM)~\cite{Gibson2021} % Chew2008
discretizes the interface between materials to reduce the size and the condition number
of matrix ${\bf{A}}$, though its matrix ${\bf{A}}$ is no longer sparse; it is efficient for homogeneous structures with a small surface-to-volume ratio.
%Doing so is optimal for homogeneous structures with small surface-to-volume ratios, though nanostructured systems typically do not have such properties.
%optical systems either have large surface-to-volume ratios due to their nanostructured constituents (as in disordered media, photonic circuits, and metasurfaces) or are not homogeneous (as in biological tissue and graded-index fibers).
%Without using the fast methods, MoM requires $\mathcal{O}(N^2)$ time to build the system matrix, $\mathcal{O}(N^3)$ time to solve, where $N$ is the number of surface elements.
Instead of a surface mesh, the $T$-matrix method~\cite{Doicu2006} uses vector spherical harmonics as basis functions, also resulting in a dense matrix ${\bf{A}}$.
The hierarchical structure of the dense matrix ${\bf{A}}$ can be utilized through the fast multipole method~\cite{Engheta1992_TAP, Song1997_TAP, Markkanen2017_JQSRT} within iterative solvers or through the $\mathcal{H}$-matrix method~\cite{Hackbusch2015} within direct solvers, but the computing time still scales as $\mathcal{O}(M)$.

% We will skip volume integral equation methods such as discrete dipole approximation (DDA)~\cite{1994_Draine_JOSAA,2007_Yurkin_JQSRT_review} because these lead to dense and large matrices, not appropriate for large systems.

For systems with a closed %({\it e.g.}, periodic)
boundary on the sides and inputs/outputs placed on the front and back surfaces, the recursive Green's function (RGF) method~\cite{Wimmer2009} can obtain the full scattering matrix without looping over the inputs,
%by slicing the system and recursively computing the side-to-side Green's function,
% with a recursion in the longitudinal direction. %MacKinnon1985_ZPB, RGF_Hsu
which is useful for disordered systems~\cite{2015_Hsu_PRL, Yilmax2019_NP}.
However, RGF works with dense Green's function matrices,
%does not utilize the sparsity ({\it i.e.}, locality) of ${\bf A}$ in the transverse directions,
so it scales unfavorably with the system width $W$ as $\mathcal{O}(W^{3(d-1)})$ for computing time and $\mathcal{O}(W^{2(d-1)})$ for memory usage in $d$ dimensions.
For layered geometries, the rigorous coupled-wave analysis (RCWA)~\cite{1996_Lalanne_JOSAA, 2014_Li_book_chapter} %, also known as the Fourier modal method~\cite{2014_Li_book_chapter},
%~\cite{Kogelnik1969_BSTJ, Moharam1981_JOSA}
and the eigenmode expansion (EME)~\cite{2001_Bienstman_thesis} methods use local eigenmodes to utilize the intralayer axial translational symmetry, which also results in dense matrices and the same scaling as RGF.

These existing methods place the emphasis on solving Maxwell's equations, typically one input at a time, after which the quantities of interest are extracted from the solutions.
Doing so is intuitive but leads to repetitions and unnecessary computations.
Here we propose an ``augmented partial factorization'' (APF) method that directly computes the entire generalized scattering matrix of interest, without dividing among the inputs and bypassing the unnecessary solution steps.
APF is general (applicable to any structure with any type of inputs and outputs, including to other linear partial differential equations), exact (no approximation beyond discretization), does not loop over the inputs, does not store large LU factors, scales well with the system size, and fully utilizes the sparsities of the Maxwell operator, of the inputs, and also of the outputs. 
These advantages lead to reduced memory usage and many-orders-of-magnitude speed-up compared to all existing methods (even the ones that specialize in a certain geometry),
enabling full-wave simulations of massively multi-channel systems that were impossible in the past. %We demonstrate so with examples.

\vspace{-4pt}
\section{\label{sec:APF}Augmented partial factorization}
\vspace{-4pt}

\noindent
Regardless of the discretization scheme (finite-difference, finite-element, boundary-element, $T$-matrix, spectral methods, {\it etc}), frequency-domain simulation for the $m$-th input reduces to computing $x_m = {\bf{A}}^{-1}b_m$, so the collective solution with the $M$ inputs is ${\bf X} = {\bf{A}}^{-1}{\bf B}$ where ${\bf{X}} = \left[ {{x_1}, \ldots ,{x_M}} \right]$ and ${\bf{B}} = \left[ {{b_1}, \ldots ,{b_M}} \right]$.
This dense and large matrix ${\bf X}$ is the conventional solution of Maxwell's equations, but its full content is rarely needed.
The needed quantities are encapsulated in the generalized scattering matrix {\bf{S}}, which we can write as
\begin{equation}
{\bf{S}} = {\bf{C}}{{\bf{A}}^{-1}}{\bf{B}} - {\bf{D}}.
\label{eq:S}
\end{equation}
Matrix ${\bf{C}}$ projects the collective solution ${\bf{X}} = {{\bf{A}}^{-1}}{\bf{B}}$ onto the $M'$ outputs of interest %through an integration
({\it e.g.}, sampling at the locations of interest, a conversion to propagating channels, or a near-to-far-field transformation);
it is sparse since the projections only use part of the solution,
and it is very fat since the number $M'$ of outputs of interest is generally far less than the number of elements Maxwell's equations are discretized onto.
Matrix ${\bf{D}} = {\bf{C}}{\bf A}_0^{-1}{\bf{B}} - {\bf S}_0$ optionally subtracts the baseline contribution, where ${\bf A}_0$ is the Maxwell operator of a reference system ({\it e.g.}, vacuum) for which the generalized scattering matrix ${\bf S}_0$ is known. 
Eq.~\eqref{eq:S} has the same appearance as scattering matrices in coupled-mode theory~\cite{2017_Alpeggiani_PRX,2020_Zhang_arXiv}, but it is exact and parameter-free. %, and requires no additional computation.

Time-dependent responses are described by a time-resolved scattering matrix, which is the Fourier transform of the frequency-domain scattering matrix~\cite{2016_Mounaix_PRL,Xiong2019_NC}.

\begin{figure*}[t]
\includegraphics[width=0.8\textwidth]{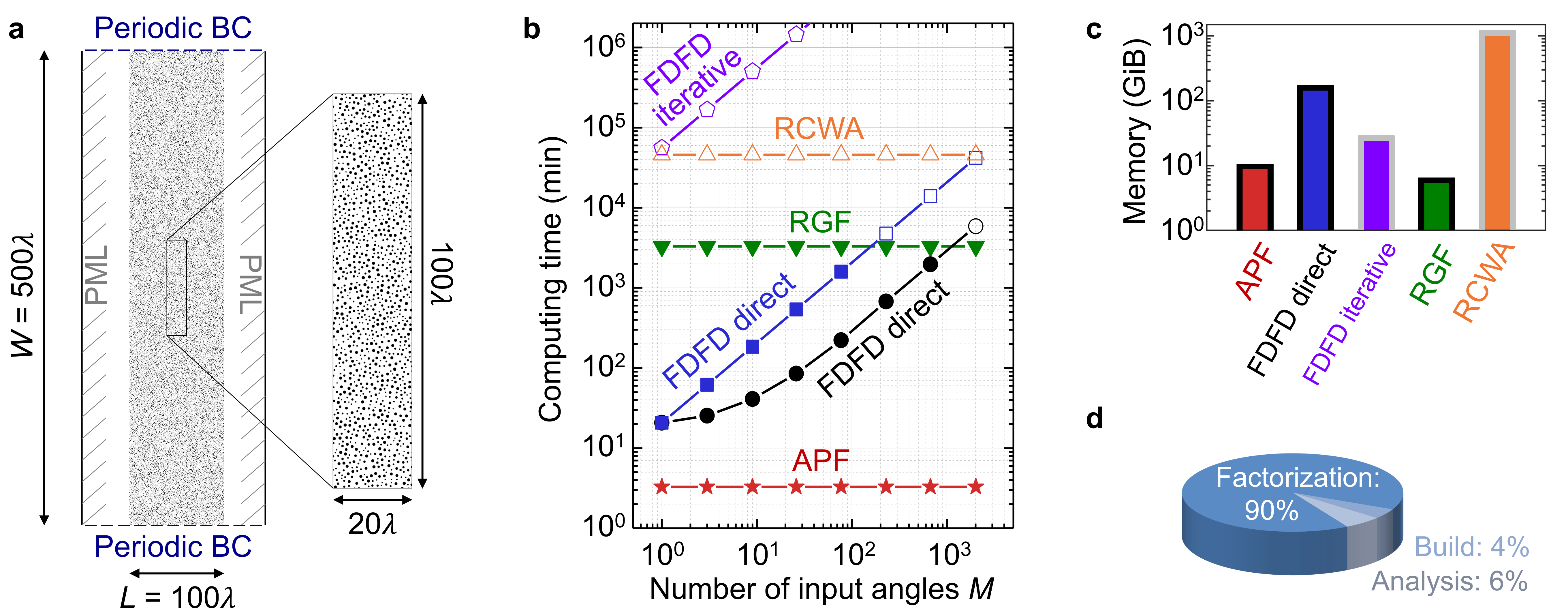}
\caption{\label{fig:disorder} {\bf Benchmarks on a large-scale disordered system.}
{\bf{a}}, Permittivity profile of the system considered; $\lambda$ is the wavelength. 
{\bf{b}}, Computing time versus the number $M$ of input angles using APF and other methods. Open symbols are extrapolated from smaller $M$ or smaller systems.
The two ``FDFD direct'' curves correspond to an unmodified version of MaxwellFDFD (blue squares) and one modified to have the LU factors reused for different inputs (black circles).
{\bf{c}}, Memory usage of different methods; gray-edged bars are extrapolated from smaller systems.
{\bf{d}}, Breakdown of the APF computing time into time used in building matrix ${\bf K}$, analyzing and reordering it, and partially factorizing it.
}
\end{figure*}

Figure~\ref{fig:schematic}c--d illustrates Eq.~\eqref{eq:S} with a concrete example. 
Consider the transverse-magnetic fields in 2D for a system periodic in $y$ with relative permittivity profile $\varepsilon_{\rm r}({\bf r}) = \varepsilon_{\rm r}(x,y)$. The Maxwell differential operator on the out-of-plane electric field $E_z({\bf r})$ at wavelength $\lambda$ is $ - {\nabla ^2} - \left( 2\pi/\lambda \right)^2{\varepsilon _{\rm{r}}}\left({\bf{r}} \right)$, which becomes matrix ${\bf A}$ when volume discretized with an outgoing boundary in $x$ direction. % realized via perfectly matched layers (PMLs)~\cite{2005_Gedney_book_chapter}.
Then matrix ${\bf A}^{-1}$ is the retarded Green's function $G({\bf r},{\bf r}')$ of this system. % that gives the response at point ${\bf r}$ given a point source at ${\bf r}'$. %, satisfying $ \left[ - {\nabla ^2} - \left( 2\pi/\lambda \right)^2{\varepsilon _{\rm{r}}}\left({\bf{r}} \right)\right]G({\bf r},{\bf r}')=\delta({\bf r}-{\bf r}')$.
An incident plane wave $e^{i (k_x^{\rm in} x+ k_y^{\rm in} y)}$ from the left can be generated with a source proportional to $\delta(x)e^{i k_y^{\rm in} y}$ on the front surface $x=0$, and incident waves from the right can be similarly generated; these source profiles become the columns of matrix ${\bf B}$ when discretized.
The coefficients of different outgoing plane waves to the left can be obtained from projections proportional to $\delta(x)e^{-i k_y^{\rm out} y}$, and similarly with outgoing waves to the right; they become the rows of matrix ${\bf{C}}$ when discretized.
%The projection matrix ${\bf{C}}$ is proportional to its conjugate transpose, ${\bf B}^{\dagger}$.
In this particular example, %matrix ${\bf{D}}$ is diagonal, and
Eq.~(\ref{eq:S}) reduces to the discrete form of the Fisher--Lee relation in quantum transport~\cite{1981_Fisher_PRB} (see Supplementary Secs.~1--2). % Datta1995, Wimmer2009
We only show a few discretized pixels and a few angles in Fig.~\ref{fig:schematic}c--d to simplify the schematic; in reality these numbers can readily exceed millions and thousands respectively.
Note that matrices ${\bf{A}}$, ${\bf{B}}$, and ${\bf{C}}$ are all sparse here.

Instead of solving for ${\bf{X}}={\bf{A}}^{-1}{\bf B}$ as is conventionally done, we directly compute the  generalized scattering matrix ${\bf{S}} = {\bf{C}}{{\bf{A}}^{-1}}{\bf{B}} - {\bf{D}}$, which is orders-of-magnitude smaller.
To do so, we build an augmented sparse matrix {\bf{K}} as illustrated in Fig.~\ref{fig:schematic}e and then perform a partial factorization,
\begin{equation}
{\bf{K}} \equiv \left[ {\begin{array}{*{20}{c}}
{\bf{A}}&{\bf{B}}\\
{\bf{C}}&{\bf{D}}
\end{array}} \right] = \left[ {\begin{array}{*{20}{c}}
{\bf{L}}&{\bf{0}}\\
{\bf{E}}&{\bf{I}}
\end{array}} \right]\left[ {\begin{array}{*{20}{c}}
{\bf{U}}&{\bf{F}}\\
{\bf{0}}&{\bf{H}}
\end{array}} \right].
\label{eq:K}
\end{equation}
The factorization is partial as it stops after factorizing the upper-left block of {\bf K} into ${\bf A}={\bf L}{\bf U}$. % with {\bf L} and {\bf U} being lower-triangular and upper-triangular, and {\bf I} being the identity matrix.
Such partial factorization can be carried out using established sparse linear-solver packages like MUMPS~\cite{Amestoy2001_SIAM} and PARDISO~\cite{2014_Petra_JSC}.
Notably, we do not use the LU factors, and {\bf L} and {\bf U} here do not even need to be triangular. By equating the middle and the right-hand side of Eq.~(\ref{eq:K}) block by block, we see that matrix {\bf H}, called the Schur complement~\cite{Zhang2005}, satisfies {\bf H} = {\bf D}$-${\bf{C}}{{\bf{A}}$^{ - 1}$}{\bf{B}}. Thus, we obtain the generalized scattering matrix via ${\bf{S}} = -{\bf{H}}$.
In this way, a single factorization yields what conventional methods obtain from $M$ separate simulations. We refer to this approach as ``augmented partial factorization'' (APF).
%The Schur complement is often used in domain decomposition~\cite{Zhang2005,Cai1999_SIAM, Dolean2009_SIAM, Smith2017,Dolean2015}, but here we use it for computing the scattering matrix without looping over the inputs.

APF is as general as Eq.~(\ref{eq:S}), applicable to any linear partial differential equation, in any dimension, under any discretization scheme, with any boundary condition, for any type of inputs generated using any scheme (such as equivalent source for arbitrary incident waves like waveguide modes~\cite{2013_Oskooi_book_chapter,Rumpf2012_PIER}, line sources, and point-dipole sources), and for any type of output projections.
It allows arbitrary material dispersion.
It is a full-wave method as precise as the underlying discretization.
%It is straightforward to implement either from ground up or based on an existing direct electromagnetic solver.

APF %solves the $M$ simulations simultaneously using a single partial factorization, avoiding 
avoids a slow loop over the $M$ inputs or a slow evaluation of the dense Green's function. 
The sparsity patterns %or hierarchical structures 
of ${\bf{A}}$, ${\bf{B}}$, and ${\bf{C}}$ are maintained in ${\bf{K}}$ and can all be utilized during the partial factorization.
Matrices {\bf{L}} and {\bf{U}} are not as sparse as ${\bf A}$, so their evaluation is slow, and their storage is the memory bottleneck for typical direct methods. 
Since APF does not compute the solution ${\bf{X}}$, such LU factors are not needed and can be dropped during the factorization. This means that APF is significantly better than conventional direct methods even when only one input ($M=1$) is considered.

\begin{figure*}[t]
\includegraphics[width=0.95\textwidth]{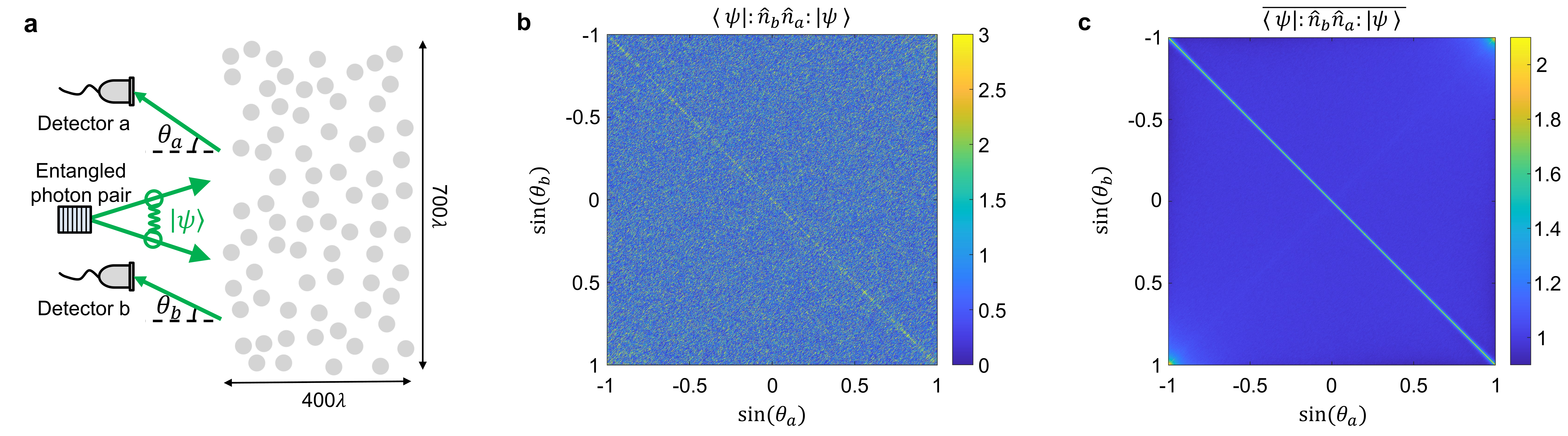}
\caption{\label{fig:2p-CBS} {\bf Two-photon coherent backscattering from disorder.}
{\bf{a}}, Schematic sketch: maximally entangled photon pairs are reflected from a dynamic disordered medium, and the photon-number correlation function ${\Gamma_{ba}} = {\langle \psi|:\hat{n}_{b}\,\hat{n}_{a}:|\psi \rangle}$ is measured for pairs of reflected angles.
{\bf{b--c}}, Normalized ${\Gamma_{ba}}$ for a single realization ({\bf b}) and averaged over 4,000 realizations ({\bf c}).
}
\end{figure*}

APF is more efficient than computing selected entries of the Green's function ${\bf A}^{-1}$~\cite{2015_Amestoy_JSC}, which does not utilize the structure of Eq.~(\ref{eq:S}).
While advanced algorithms have been developed to exploit the sparsity of the input and the output during forward and backward substitutions~\cite{2019_Amestoy_CG} or through domain decomposition~\cite{2012_Hackbusch_CVS}, they still require an $\mathcal{O}(M)$ substitution stage, with a modest speed-up (a factor of 3 when $M$ is several thousands) and no memory usage reduction.
APF is simpler yet much more efficient as it completely obviates the forward and backward substitution steps and the need for LU factors.

Typically, matrix ${\bf{A}}$ contains more nonzero elements than matrices ${\bf{B}}$ and ${\bf{C}}$, and we find in such scenarios that the computing time and memory usage of APF scale as $\mathcal{O}(N^{1.3})$ and $\mathcal{O}(N)$ respectively in 2D (Supplementary Fig.~S1), where
$N$ is the number of nonzero elements in matrix ${\bf{K}}$ and is almost independent of $M$. %; it is proportional to the system size and almost independent of the number $M$ of input channels. 

When ${\bf{B}}$ and/or ${\bf{C}}$ contain more nonzero elements than ${\bf{A}}$, % (such as the metasurface examples in Sec.~\ref{sec:metalens}),
we can compress matrices {\bf{B}} and {\bf{C}} through a data-sparse representation to reduce their numbers of nonzero elements to below that of ${\bf A}$. For example, a plane-wave source spans a large area, but a superposition of plane waves can be spatially localized and truncated with negligible error (Supplementary Sec.~5 and Figs.~S2--S3).

We implement APF under finite-difference discretization on the Yee grid (Supplementary Secs.~2--3), with outgoing boundaries realized by perfectly matched layer (PML)~\cite{2005_Gedney_book_chapter}.
%and using the MUMPS package~\cite{Amestoy2001_SIAM} with AMD ordering~\cite{Amestoy1996_SIAM} to compute the Schur complement.
%Since MUMPS only works with square matrices, zeros are padded to ${\bf B}$ or ${\bf C}$ when $M' \neq M$.
%We order the output channels and/or pad additional channels so that matrix ${\bf K}$ is symmetric (Supplementary Sec.~3).
Pseudocodes are shown in Supplementary Sec.~6.
Our code supports all common boundary conditions and allows arbitrary permittivity profiles and arbitrary inputs and outputs (user-specified or automatically built); we have made it open-source~\cite{MESTI_GitHub}.

Below, we consider two classes of massively multi-channel systems, %disordered media and aperiodic metasurfaces,
comparing the computing time, memory usage, and accuracy of APF to widely-used open-source electromagnetic solvers including a conventional finite-difference frequency-domain (FDFD) code MaxwellFDFD using either (1) direct~\cite{MaxwellFDFD} or (2) iterative~\cite{FD3D} methods, % used in the photonics inverse design package SPINS.~\cite{Su2020_APR}
(3) an RGF code~\cite{RGF_Hsu}, and (4) an RCWA code S4~\cite{Liu2012_CPC}; see the Methods section for details.
Time-domain methods typically require more iterations than their iterative frequency-domain counterpart~\cite{2016_Osnabrugge_JCP} since they iterate by time stepping;
we do not include time-domain methods here since the comparisons below are made on monochromatic responses.
All computations are done in serial on identical processors.
We consider transverse-magnetic fields in 2D, starting with systems small enough for these solvers,
%, even though APF can handle larger ones.
then with larger problems that only APF can tackle. %: coherent backscattering of entangled photon pairs from disordered media, and the angle-dependent response of millimeter-wide optical metalenses.

\vspace{-4pt}
\section{\label{sec:disordered_media} Large-scale disordered systems}
\vspace{-4pt}

\noindent
Disordered systems are difficult to simulate given their large size-to-wavelength ratio, large number of channels, strong scattering, and lack of symmetry.
Here we consider one $W=500\lambda$ wide and $L=100\lambda$ thick, consisting of 30,000 cylindrical scatterers with refractive index of 2.0 in air (Fig.~\ref{fig:disorder}a), discretized %with subpixel smoothing~\cite{2006_Farjadpour_OL}
into over 11 million pixels % with grid size $\lambda/15$.
with a periodic boundary condition in $y$ to mimic an infinitely-wide system.
On each of the $-x$ and $+x$ sides, $2W/\lambda =$ 1,000 channels (plane waves with different angles) are necessary to specify the propagating components of an incident wavefront or outgoing wavefront, which is the minimal requirement at the Nyquist sampling rate
(Supplementary Sec.~1A).
So, we compute the scattering matrix with $M'=$ 2,000 outputs and up to $M=$ 2,000 inputs (including both sides).

It takes APF 3.3 minutes and 10 GiB of memory to compute the full scattering matrix; the other methods take between 3,300 and 110,000,000 minutes using between 7.0 to 1,200 GiB of memory for the same computation (Fig.~\ref{fig:disorder}b--c).
The computing times of APF (with breakdown shown in Fig.~\ref{fig:disorder}d), RGF, and RCWA are all independent of $M$, though APF is orders-of-magnitude faster.
MaxwellFDFD takes $\mathcal{O}(M)$ time due to its loop over the inputs; reusing the LU factors helps, but the $M$ forward and backward substitutions become the bottleneck when $M \gtrsim 20$.
Note that APF offers substantial advantage even in the single-input ($M=1$) case.

\begin{figure*}
\includegraphics[width=0.75\textwidth]{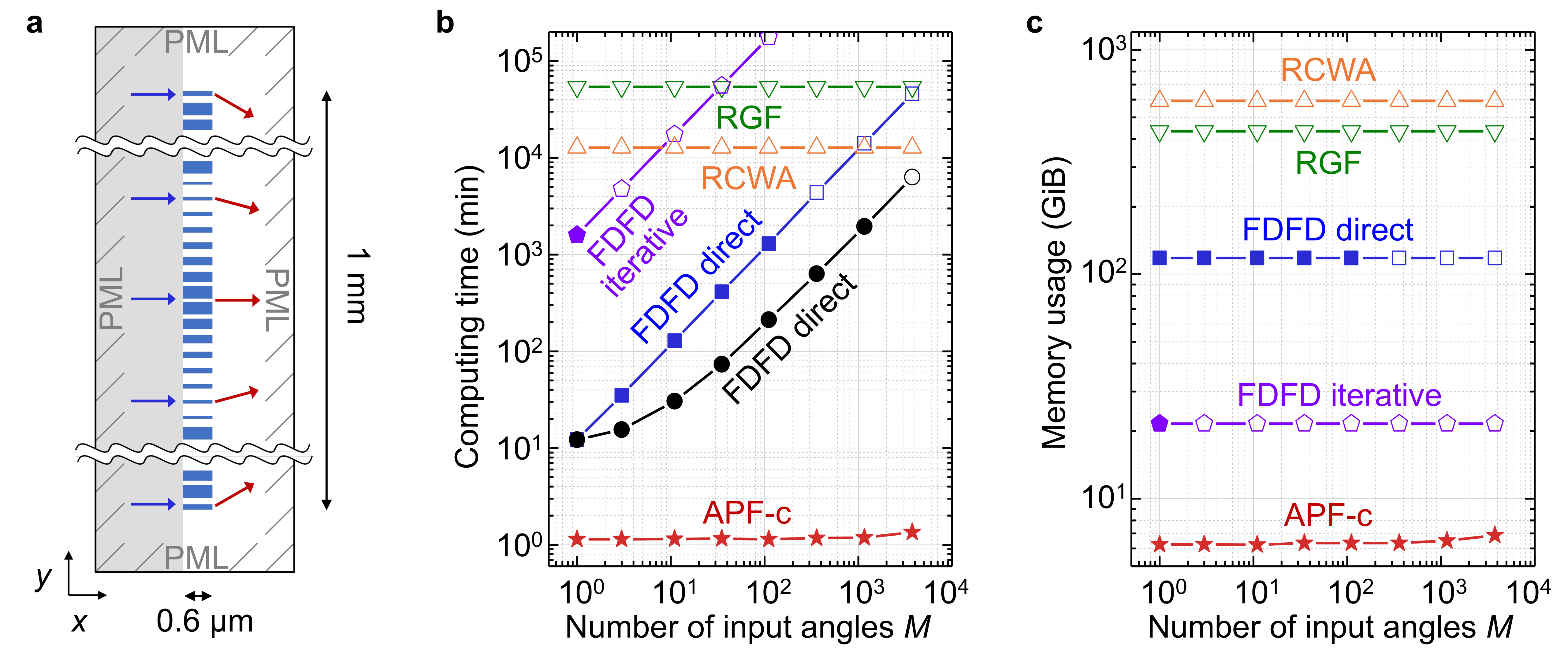}
\caption{\label{fig:meta1} {\bf Benchmarks on a large-area metasurface.}
{\bf{a}}, Schematic of the system: a mm-wide metasurface consisting of 4,178 titanium dioxide ridges (blue) on a silica substrate (gray), operating at wavelength $\lambda = 532$ nm.
{\bf{b}--c}, Computing time and memory usage versus the number $M$ of input angles using different methods. Open symbols are extrapolated from smaller $M$ or smaller systems.  APF-c denotes APF with matrices ${\bf B}$ and ${\bf C}$ compressed.
}
\end{figure*}

The speed and memory advantage of APF further grows with the system size (Supplementary Fig.~S4).
Some of these solvers require more computing resources than we have access to, so their usage data (open symbols and gray-edged bars in Fig.~\ref{fig:disorder}b--c) are extrapolated based on smaller systems (Fig.~S4).

The relative $\ell^2$-norm error of APF due to numerical round-off %(its only error beyond discretization)
is $10^{-12}$ here and grows slowly with an $\mathcal{O}(N^{1/2})$ scaling (Supplementary Fig.~S5), while the iterative MaxwellFDFD here has a relative $\ell^2$ error of $10^{-6}$.

In Ref.~\citen{Safadi2022_ArXiv}, it is predicted that entangled photon pairs remain partially correlated even after multiple scattering from a dynamic disordered medium. We demonstrate such ``two-photon coherent backscattering'' as an example here.
Given a maximally entangled input state, the correlation between two photons reflected into directions $\theta_a$ and $\theta_b$ is %~\cite{Safadi2022_ArXiv}
\begin{equation} \label{eq:Gamma}
\overline{\Gamma_{ba}} = \overline{\langle \psi|:\hat{n}_{b}\,\hat{n}_{a}:|\psi \rangle}
\propto \overline{\left|(r^2)_{\theta_b,-\theta_{a}}\right|^2},
\end{equation}
where $|\psi\rangle$ is the two-photon wave function, $\hat{n}_{a}$ is the photon number operator in reflected direction $\theta_{a}$, $:(\dots):$ stands for normal ordering, $r^2$ is the square of the medium's reflection matrix ({\it i.e.}, the scattering matrix with inputs and outputs on the same side), and $\overline{\mbox{\dots\rule{0pt}{1.5mm}}}$ indicates ensemble average over disorder realizations.
This requires the full reflection matrix with all incident angles and all outgoing angles, for many realizations, and the disordered medium must be wide (for angular resolution) and thick (to reach diffusive transport).
Figure~\ref{fig:2p-CBS} shows the two-photon correlation function $\Gamma_{ba}$ computed with APF before and after averaging over 4,000 disorder realizations, for a system $W=700\lambda$ wide and $L=400\lambda$ thick consisting of 56,000 cylindrical scatterers with a transport mean free path of $\ell_{\rm t} = 9.5\lambda$.
We find the correlation between photons reflected toward similar directions ($|\theta_b - \theta_a| \lesssim 0.1 \lambda/\ell_t$) to be enhanced by a factor of 2.
This finding, described in detail in Ref.~\citen{Safadi2022_ArXiv}, is the first evidence of two-photon coherent backscattering in disordered media.
These APF simulations use modest resources on a computing cluster but would have taken prohibitively long using any other method.
%Simulating such two-photon coherent backscattering will be at least 1,000 times slower using any method other than APF.

\vspace{-4pt}
\section{\label{sec:metalens} Large-area metasurfaces}
\vspace{-4pt}

\noindent
Metalenses are lenses made with metasurfaces~\cite{lalanne2017metalenses, Khorasaninejad2017_Sci}.
When the numerical aperture (NA) is high, metalenses need to generate large phase gradients, so the variation from one unit cell to the next must be large, and the locally periodic approximation (LPA)~\cite{Pestourie2018_OE,Kamali2018_Nanophotonics} fails.
Full-wave simulation remains the gold standard.
Here we consider metalenses with height $L=0.6$ {\textmu}m and width $W \approx 1$ mm, consisting of 4,178 unit cells of titanium dioxide ridges on a silica substrate (Fig.~\ref{fig:meta1}a),
for a hyperbolic~\cite{aieta2012aberration} phase profile with NA $= 0.86$ and a quadratic~\cite{Pu2017_OE, Martins2020_ACSP, Lassalle2021_ACSP} phase profile with NA $=0.71$ operating
at wavelength $\lambda = 532$ nm;
see Supplementary Sec.~9 and Fig.~S6 for design and simulation details.
PMLs are placed on all sides, and the system is discretized with grid size $\Delta x = \lambda/40$ 
into over 11 million pixels.
We compute the transmission matrix at the minimal Nyquist sampling rate, with up to $M=2W/\lambda = $ 3,761 plane-wave inputs from the left (substrate side) truncated within the width $W$ of the metalens, and sampling the transmitted field across width $W_{\rm out} = W + 40\lambda$ (to ensure all transmitted light is captured) projected onto $M'=2W_{\rm out}/\lambda =$ 3,841 propagating plane waves on the right.
Due to the large (1 mm)/(0.6 {\textmu}m) aspect ratio, the number of nonzero elements in matrices {\bf{B}} and {\bf{C}} is larger than that of {\bf{A}}, so we compress {\bf{B}} and {\bf{C}} and denote this as APF-c (Supplementary Sec.~5).

\begin{figure*}
\includegraphics[width=0.85\textwidth]{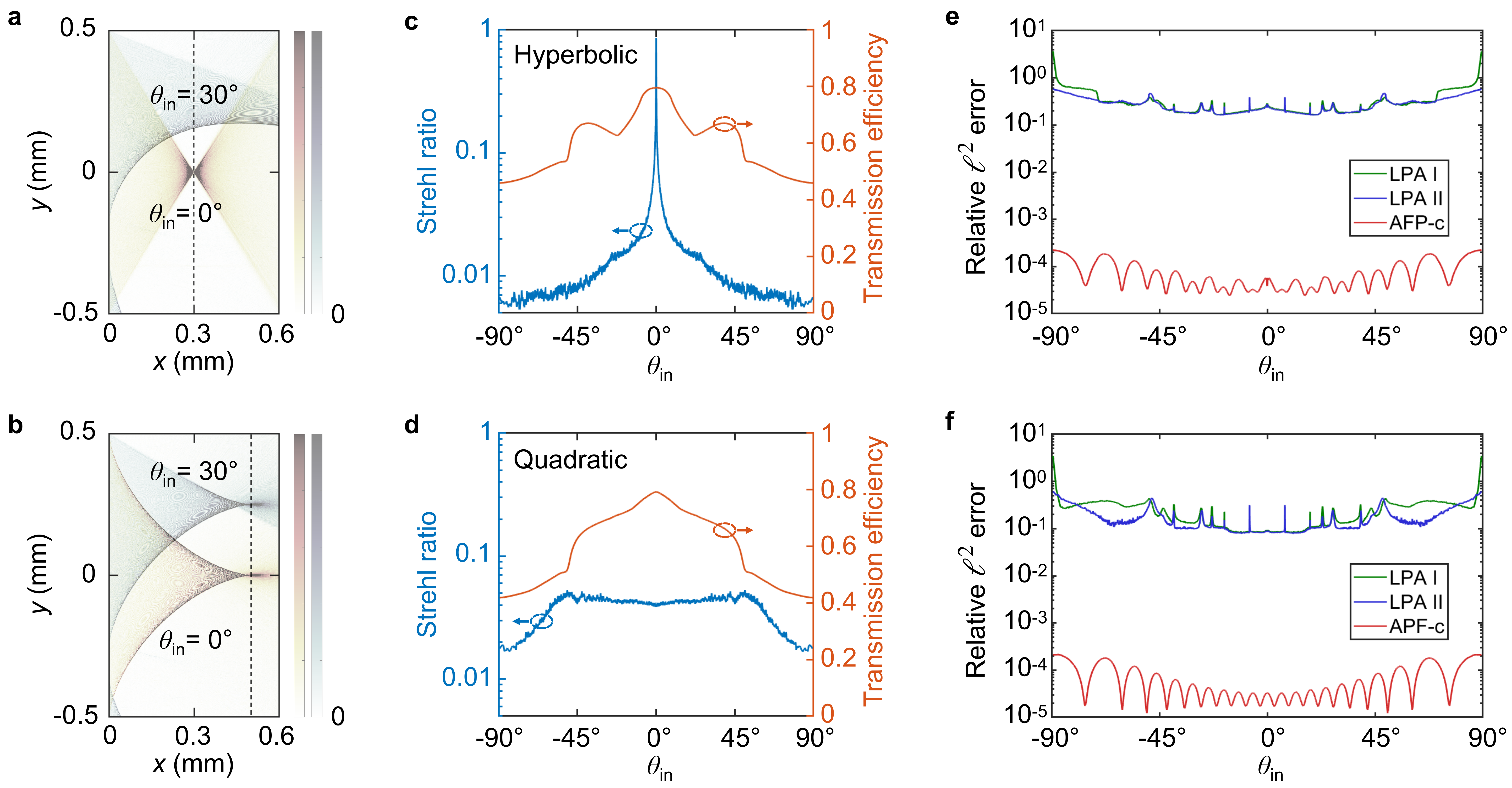}
\caption{\label{fig:meta2} {\bf All-angle full-wave characterization of mm-wide metalenses.} 
{\bf{a--b}}, Saturated intensity profiles of the transmitted light for two incident angles with a hyperbolic metalens ({\bf a}) and a quadratic metalens ({\bf b}). Black dashed lines indicate the focal plane. $\theta_{\rm in} = \sin^{-1}(n_{\rm substrate}\sin\theta_{\rm in}^{\rm substrate})$ is the incident angle in air. 
Complete dependence on $\theta_{\rm in}$ is shown in Supplementary Movies S1--S2.
{\bf{c--d}}, 
Angle dependence of the Strehl ratio and transmission efficiency.
{\bf{e--f}}, Errors of locally periodic approximation (LPA) and of APF-c.
}
\end{figure*}

It takes APF-c 1.3 minute and 6.9 GiB of memory to compute this transmission matrix, while the other methods take 6,300 to 6,000,000 minutes using 22 to 600 GiB (Fig.~\ref{fig:meta1}b--c); some numbers are extrapolated from smaller systems (Supplementary Fig.~S7).
It is remarkable that even though RCWA is specialized for layered structures like the metasurface here, the general-purpose APF-c still outperforms RCWA by 10,000 times in speed and 87 times in memory.
The second-best solver here is MaxwellFDFD with the LU factors stored and reused, which takes 4,800 times longer while using 18 times more memory compared to APF-c.

The transmission matrix fully characterizes the metasurface's response to any input.
Here we use it with angular spectrum propagation (Supplementary Sec.~11) to obtain the angle dependence of the exact transmitted profile (two profiles each shown in Fig.~\ref{fig:meta2}a--b; more shown in Supplementary Movies S1--S2), the Strehl ratio, and the transmission efficiency (Fig.~\ref{fig:meta2}c--d and Supplementary Sec.~12).
They rigorously quantify the performance of these metalens designs.
To our knowledge, this is the first time full-wave simulations over all angles are carried out for a large-area metasurface.

To quantify the accuracy of an approximation, we compute the relative $\ell^2$-norm error ${\left\| I - I_0 \right\|_2}/{\left\| I_0 \right\|_2}$, with $I_0$ being the intensity at the focal plane within $|y|<W/2$ calculated from APF without compression, and $I$ from an approximation.
We consider two LPA formalisms, a standard one using the unit cells' propagating fields (LPA I) and one with the unit cells' evanescent fields included (LPA II); see Supplementary Sec.~13.
LPA leads to errors ranging between $8\%$ and 366$\%$ depending on the incident angle, with the angle-averaged error between $18\%$ and $37\%$ (Fig.~\ref{fig:meta2}e--f).
Meanwhile, the compression errors of APF-c here average below $0.01\%$ (Fig.~\ref{fig:meta2}e--f) and can be made arbitrarily small (Supplementary Fig.~S8). 

\vspace{-4pt}
%\section{\label{sec:outlook}Outlook}
\section{\label{sec:outlook}Discussion}
\vspace{-4pt}

\noindent
The APF method opens the door to a wide range of studies. 
The correlation functions of multi-photon scattering from disorder are challenging to measure experimentally but can now be computed exactly with APF.
With volumetric imaging inside strongly scattering media such as biological tissue~\cite{Jin2017_JOptSocAmB,Park2018_NP,Yoon2020_NRP,Badon2020_SA}, it is difficult to experimentally validate against a ground truth and to separate factors like aberrations and dispersion from scattering, but large-scale simulations enabling such studies are now possible with APF~\cite{2022_Wang_arXiv}.
Inverse design using the adjoint method used to require $2M$ simulations given $M$ inputs~\cite{Lin2021_APL}, limiting the number of inputs that can be considered;
with a suitable formulation, APF can consolidate the $2M$ simulations into a single computation.
It may now be possible, for example, to design broad-angle nonlocal metasurfaces that reach fundamental bounds~\cite{Li2022_ArXiv}.
Predicting the thermal emission from nanostructures~\cite{2021_Zhou_ACSphoton_1} requires a large number of frequency-domain simulations with different point-dipole sources, which can be vastly accelerated using APF.
Classical~\cite{2017_Shen_nphoton, 2020_Wetzstein_Nature_perspective} and quantum~\cite{2021_Pelucchi_NRP,2022_Madsen_Nature} photonic circuits are currently built from small components that couple very few channels at a time, and APF can enlarge the design space to large multi-channel elements with more functionality while occupying a compact footprint.

Beyond photonics, APF can be used for mapping the angle dependence of radar cross-sections, for microwave imaging~\cite{2014_Haynes_TBE}, for full waveform inversion~\cite{2009_Virieux_Geophysics_review} and controlled-source electromagnetic surveys~\cite{2019_Amestoy_CG} in geophysics, and for quantum transport simulations~\cite{2014_Groth_NJP}.
More generally, APF can efficiently evaluate matrices of the form ${\bf C}{\bf A}^{-1}{\bf B}$ for other applications of numerical linear algebra, not limited to wave physics or to partial differential equations.

As the number of channels and the LU factor size are both much larger in 3D, the advantage of APF over existing methods will be even more significant in 3D than in 2D. %, for both computing time and memory.
With the low-rank property of the finite-difference matrix ${\bf A}$ utilized, APF should take $\mathcal{O}(N^{4/3})$ time in 3D following the known factorization cost~\cite{2017_Amestoy_JSC}, which is comparable to the 2D scaling (Supplementary Fig.~S1).
APF-c can naturally be used with overlapping-domain distribution strategies~\cite{Lin2019_OE,Torfeh2020_ACSP,2022_Skarda_CM} % 2019_Phan_LSA
when modeling systems with large aspect ratios.
Multi-frontal parallelization can be used through existing packages such as MUMPS~\cite{Amestoy2001_SIAM}, and one may employ hardware accelerations with GPUs~\cite{Egel2017JQSRT, 2022_Skarda_CM, Hughes2021_APL}. 
For systems with a small surface-to-volume ratio, it is also possible to apply APF to BEM or $T$-matrix method, using the $\mathcal{H}$-matrix technique~\cite{Hackbusch2015} for fast factorization.
There is room to further reduce the computing cost of ${\bf C}{\bf A}^{-1}{\bf B}$ below that of existing factorization algorithms since the LU factors are not needed and do not need to be triangular.
This elimination of LU factors makes APF favorable over iterative methods even for single-input scenarios.

\vspace{-4pt}
\begin{acknowledgments}
\vspace{-4pt}
{\small
We thank Y.~Bromberg, M.~Safadi, A.~Goetschy, C.~Sideris, M.~Povinelli, A.~D.~Stone, S.~Li, M.~Torfeh, and Y.~Zhang for useful discussions.
{\bf Funding:} This work is supported by the National Science Foundation CAREER award (ECCS-2146021) and the Sony Research Award Program.
Computing resources are provided by the Center for Advanced Research Computing at the University of Southern California.
{\bf Author contributions:} 
H.-C.L.~and C.W.H.~performed the simulations and data analysis; C.W.H., H.-C.L., and Z.W.~wrote the APF codes;
C.W.H.~developed the theory and supervised research;
all authors contributed to designing the systems, discussing the results, and preparing the manuscript.
{\bf Competing interests:} The authors declare no competing interests.
{\bf Data availability:} All data needed to evaluate the conclusions in this study are presented in the paper and in the supplementary materials. 
{\bf Code availability:} Source codes are available at Ref.~\citen{MESTI_GitHub}.
}
\end{acknowledgments}

\bibliographystyle{naturemag}
\bibliography{main}% Produces the bibliography via BibTeX.

%\bigskip
%\vfill\eject % force column break
%\bigskip
%\smallskip

\vspace{-4pt}
\section*{Methods}
\vspace{-4pt}

{\small
We use the MUMPS package~\cite{Amestoy2001_SIAM} with the default AMD ordering~\cite{Amestoy1996_SIAM} to compute the Schur complement with partial factorization.
%Since MUMPS only works with square matrices, zeros are padded to ${\bf B}$ or ${\bf C}$ when $M' \neq M$.
We order the output channels and/or pad additional channels so that matrix ${\bf K}$ is symmetric (Supplementary Sec.~3).

We use the same second-order finite-difference discretization scheme (Supplementary Sec.~2), same grid size, and same subpixel smoothing~\cite{2006_Farjadpour_OL} for the APF, MaxwellFDFD, and RGF benchmarks. 
Discretization error is not important for the disordered media example in Fig.~2, so we use a relatively coarse resolution of 15 pixels per $\lambda$ there. A finer resolution of 40 pixels per $\lambda = 532$ nm is used for the metasurface examples in Figs.~4--5 to have their transmission phase shifts accurate to within 0.1 radian (Supplementary Fig.~S9).

In RGF~\cite{RGF_Hsu}, the outgoing boundary in the longitudinal direction is implemented exactly through the retarded Green's function of a semi-infinite discrete space~\cite{Wimmer2009}.
For APF and MaxwellFDFD, one $\lambda$ of homogeneous space and 10 pixels of PML~\cite{2005_Gedney_book_chapter} are used to achieve an outgoing boundary with a sufficiently small discretization-induced reflection~\cite{1996_Chew_electromagnetics}.
Uniaxial PML is used in APF so that matrix ${\bf A}$ is symmetric. Stretched-coordinate PML is used in MaxwellFDFD to lower the condition number~\cite{Shin2012_JCP}.

For MaxwellFDFD with an iterative solver~\cite{FD3D}, we use its default biconjugate gradient method with its default convergence criterion of relative $\ell^2$ residual below $10^{-6}$.
For MaxwellFDFD with a direct solver~\cite{MaxwellFDFD}, we consider an unmodified version where the LU factors are not reused and a version modified to have the LU factors stored in the memory and reused for the different inputs.

For the RCWA simulations, we use its default closed-form Fourier transform formalism implemented in S4~\cite{Liu2012_CPC}.
For the example in Fig.~4, we use a single layer with 5 Fourier components per unit cell where the cell width is 239 nm ({\it i.e.}, 11 Fourier components per $\lambda$), which gives comparable accuracy as APF, MaxwellFDFD, and RGF (Supplementary Fig.~S9).
For the example in Fig.~2, we use 15 layers per $\lambda$ axially (same as the discretization grid size used in the other methods) with 4.1 Fourier components per $\lambda$ laterally (by scaling it down in proportion with the reduced spatial resolution in APF, MaxwellFDFD, and RGF).

Note that the RGF~\cite{RGF_Hsu} and S4~\cite{Liu2012_CPC} codes do not support an outgoing boundary in the transverse $y$ direction. 
The RGF and S4 computing time and memory usage in Fig.~4 are extrapolated from simulations on smaller systems adopting a periodic transverse boundary (Supplementary Fig.~S7).
To simulate the example in Fig.~4 using RGF or S4, one needs to additionally implement PML in the $y$ direction (as in Refs.~\citen{2002_Bienstman_OQE, 2005_Hugonin_JOSAA}) and to further increase the system width;
doing so will slightly increase their computing time and memory usage, which we do not account for.

All timing and memory usage numbers are obtained from serial computations on identical Intel Xeon Gold 6130 nodes on the USC Center for Advanced Research Computing's Discovery cluster with 191 GiB of memory per node.
}

\end{document}

\title{Supplementary Materials\\
%Full-wave solver for massively multi-channel optics using augmented partial factorization
%Full-wave multi-source nanophotonic simulations without solving Maxwell's equations
Fast multi-source nanophotonic simulations using augmented partial factorization
}
\author{Ho-Chun Lin}
\author{Zeyu Wang}
\author{Chia Wei Hsu}
%\email{cwhsu@usc.edu}
\affiliation{%
Ming Hsieh Department of Electrical and Computer Engineering, University of Southern California, Los Angeles, California 90089, USA
}%

\maketitle

\tableofcontents

\section{Scattering matrix \label{sec:s_matrix}} 

In this section, we define the flux-normalized scattering matrix and derive its expression. For concreteness, here we consider the 2D transverse-magnetic fields in a two-sided system, with outgoing boundary condition in the longitudinal direction $x$ and Bloch periodic boundary condition in the transverse direction $y$. Defining the corresponding expressions for other boundary conditions and in 3D is straightforward, and the concepts apply similarly to other geometries.

We start with time-harmonic electromagnetic waves at frequency $\omega$ with $e^{-i\omega t}$ dependence on time $t$, governed by the frequency-domain Maxwell's equations,
\begin{subequations}
\label{eq:Maxwell_3D}
\begin{align}
\label{eq:faraday}
\nabla  \times {\bf{E}}({\bf{r}}) - i\omega {\mu _0}{\mu _{\rm{r}}}({\bf{r}}){\bf{H}}( {\bf{r}}) &= 0,\\
\label{eq:ampere}
\nabla  \times {\bf{H}}({\bf{r}}) + i\omega {\varepsilon _0}{\varepsilon _{\rm{r}}}( {\bf{r}}){\bf{E}}({\bf{r}}) &= {\bf{J}}({\bf{r}}),\\
\label{eq:gauss}
\nabla  \cdot \left[ {{\varepsilon _{\rm{r}}}({\bf{r}}){\bf{E}}({\bf{r}})} \right] &= \rho({\bf{r}})/{\varepsilon _0},\\
\label{eq:gauss_m}
\nabla  \cdot \left[ {{\mu _{\rm{r}}}({\bf{r}}){\bf{H}}({\bf{r}})} \right] &= 0,
\end{align} 
\end{subequations}
where ${\bf E}({\bf r})=(E_x,E_y,E_z)({\bf r})$ and ${\bf H}({\bf r})=(H_x,H_y,H_z)({\bf r})$ are the electric and magnetic fields at position ${\bf r}=(x,y,z)$, $\varepsilon_{\rm r}({\bf r})$ and $\mu_{\rm r}({\bf r})$ the scalar relative permittivity and permeability profiles that define the structure, ${\bf J}({\bf r})$ and $\rho({\bf r})$ the electric current density and charge density satisfying continuity equation $\nabla \cdot {\bf{J}}({\bf{r}}) = i\omega \rho({\bf r})$ that act as the sources, and $\varepsilon_0$ and $\mu_0$ the vacuum permittivity and permeability constants.

Consider a nonmagnetic (${\mu _{\rm{r}}} = 1$) system where $\varepsilon_{\rm r}({\bf r})$, ${\bf E}({\bf r})$, ${\bf H}({\bf r})$, ${\bf J}({\bf r})$, and $\rho({\bf r})$ are independent of $z$. Then, the transverse-magnetic (TM) field components $H_x$, $H_y$, and $E_z$ are decoupled from the transverse-electric (TE) components $E_x$, $E_y$, and $H_z$. 
Inserting the ${\bf H}({\bf r})$ in Eq.~\eqref{eq:faraday} into Eq.~\eqref{eq:ampere} and taking the $z$ component, we obtain the governing equation for the TM waves,
\begin{equation}
\label{eq:A_psi}
\left[ -\nabla_{xy}^2 - {\frac{\omega^2 }{c^2}} \varepsilon_{\rm{r}}(x,y)\right] E_z(x,y) = i \omega \mu_0 J_z(x,y),
\end{equation}
where $\nabla_{xy}^2 = \partial^2/\partial x^2 + \partial^2/\partial y^2$, and $c=1/\sqrt{\varepsilon_0 \mu_0}$ is the vacuum speed of light.  Once $E_z$ is solved for, the magnetic-field components of the TM wave follow from Eq.~\eqref{eq:faraday} as $\left( {{H_x},{H_y}} \right) = \frac{1}{{i\omega {\mu _0}}}\left( {\frac{{\partial {E_z}}}{{\partial y}},-\frac{{\partial {E_z}}}{{\partial x}}} \right)$. Eqs.~\eqref{eq:gauss}--\eqref{eq:gauss_m} are automatically satisfied. %Eq.~\eqref{eq:gauss} is also automatically satisfied since $\partial E_z/\partial z = 0$.
The vacuum wavelength is $\lambda = 2\pi c/\omega$. 

For the scattering problems considered below, there is no physical current source, so the right-hand side $J_z$ is zero. We will put effective sources on the right-hand side of Eq.~\eqref{eq:A_psi} to generate the incident field and to solve for the resulting scattered field, but the total field $E_z$ still satisfies Eq.~\eqref{eq:A_psi} with $J_z=0$; see Sec.~\ref{sec:Fisher_Lee}.

Consider a structure where ${\varepsilon _{\rm{r}}}(x,y)$ is periodic in $y$ with periodicity $W$, namely ${\varepsilon _{\rm{r}}}(x,y+W) = {\varepsilon _{\rm{r}}}(x,y)$.
Given the periodicity, any solution $E_z(x,y)$ of Eq.~\eqref{eq:A_psi} can be written as a superposition of Bloch states with different Bloch wave numbers $k_{\rm B}$, with each Bloch state satisfying $E_z(x,y + W) = E_z(x,y){e^{i{k_{\rm{B}}}W}}$.
Bloch states with distinct $k_{\rm B}$ (up to modulo $2\pi/W$) are decoupled from each other.
Therefore, we can fix $k_{\rm B}$ and consider only a finite transverse size within $0 \le y \le W$, with boundary condition $E_z(x,W) = E_z(x,0) e^{i k_{\rm B} W}$ at $y=0$ and $y=W$; this is called Bloch periodic boundary condition, which reduces to a periodic boundary when $k_{\rm B}=0$.

We further consider the structure to be homogeneous on the left and right sides,
\begin{equation}
\label{eq:syst}
{\varepsilon _{\rm{r}}}(x,y) =
\begin{cases}
\, \varepsilon_{\rm L}, & x \le 0, \\
\, \varepsilon_{\rm{r}}(x,y), & 0 < x < L, \\
\, \varepsilon_{\rm R}, & x \ge L.
\end{cases}
\end{equation}
Light incident from either side is scattered by the inhomogeneous structure within $0 < x < L$, which we refer to as the scattering region. 
The scattering matrix fully characterizes such response.
To define the scattering matrix, we first define the input and output ``channels'' as follows.

\subsection{Channels in homogeneous space\label{sec:chan_homo_sp}}
Consider a homogeneous region (either $x\le0$ or $x\ge L$), within which
$\varepsilon_{\rm r}({x,y}) = {\varepsilon _{{\rm{bg}}}}$ is a real-valued constant (either $\varepsilon_{\rm L}$ or $\varepsilon_{\rm R}$).
Given the translational symmetry in $x$, the field profile within the homogeneous region can be written as a superposition of propagating and evanescent fields of the form
\begin{equation}
\label{eq:Ez_a}
E_{z}^{(a,\pm)}(x,y) = \frac{1}{\sqrt{k_x^{(a)}}}  u_a(y) e^{ \pm i k_x^{(a)} x }, \quad a \in \mathbb{Z},
\end{equation} 
which we refer to as ``channels.'' The integer index $a$ labels the channel number, $\pm$ indicates the propagation direction, $k_x^{(a)}$ is the longitudinal wave number, the prefactor $1/\sqrt{k_x^{(a)}}$ normalizes the longitudinal flux (to be discussed in the next paragraph), and $u_a(y)$ is the transverse mode profile. The $a$-th transverse profile is \begin{equation}
\label{eq:phi_a}
u_a(y) = \frac{1}{{\sqrt W }}\exp \left[ {ik_y^{\left( a \right)}\left( {y - {y_0}} \right)} \right], \quad 0 \le y \le W,
\end{equation} 
with the transverse wave number $k_y^{\left( a \right)} = {k_{\rm{B}}} + a\frac{{2\pi }}{W}$ chosen to satisfy the Bloch periodic boundary condition; note that neighboring $k_y^{\left( a \right)}$ are separated by $2\pi/W$.
Here, $y_0$ is an arbitrary real constant specifying the reference position of the basis.
The set $\{ u_a(y) \}_a$ of transverse modes makes up a complete and orthonormal basis, with $\int_0^W dy u_a^*(y)u_b(y) = \delta_{ab}$.
Inserting Eqs.~\eqref{eq:Ez_a}--\eqref{eq:phi_a} into Eq.~\eqref{eq:A_psi} with no source yields the dispersion relation
$\left( {{\omega }/{c}} \right)^2{\varepsilon _{{\rm{bg}}}} = (k_x^{(a)})^2 + (k_y^{(a)})^2$.
We choose the sign of $k_x^{(a)}$ as $k_x^{(a)} = \sqrt{(\omega/c)^2\varepsilon_{\rm bg} - (k_y^{(a)})^2}$.
There is an infinite number of evanescent channels where $k_x^{(a)}$ is imaginary [{\it i.e.}, where $|k_y^{\left( a \right)}| > (\omega/c)\sqrt{\varepsilon _{\rm{bg}}}$].
There are approximately $2\sqrt{\varepsilon _{\rm{bg}}}W/\lambda$ propagating channels where $k_x^{(a)}$ is real-valued [{\it i.e.}, where $|k_y^{\left( a \right)}| < (\omega/c)\sqrt{\varepsilon _{\rm{bg}}}$]; these are plane waves propagating at angles $\theta_a = {\rm atan}(k_y^{(a)}/k_x^{(a)})$.
Therefore, $2\sqrt{\varepsilon _{\rm{bg}}}W/\lambda$ complex-valued coefficients are necessary to fully specify the propagating components of an arbitrary incident wavefront or outgoing wavefront;
this is consistent with the Nyquist--Shannon sampling theorem~\cite{landau1967sampling} that when the highest spatial frequency ({\it i.e.}, wave number $k_y$) in a wavefront is $(\omega/c)\sqrt{\varepsilon _{\rm{bg}}}$, we need at least one spatial sampling point per $\lambda/(2\sqrt{\varepsilon _{\rm{bg}}})$ to uniquely specify such a wavefront.

The channels are flux orthogonal:
given a superposition of Eq.~\eqref{eq:Ez_a} with different channel indices and propagation directions,
the $x$-directional Poynting flux of the total field integrated over $y$ equals the sum of the integrated Poynting flux for components with different transverse profiles.
For propagating channels, the $1/\sqrt{k_x^{(a)}}$ prefactor in Eq.~\eqref{eq:Ez_a} ensures that different propagating channels are normalized to carry the same longitudinal flux; furthermore, the integrated flux from the two counter-propagating terms have opposite signs, and the cross term does not contribute.
For evanescent channels, neither $E_{z}^{(a,+)}$ nor $E_{z}^{(a,-)}$ itself carries flux, but their cross terms can carry flux.

\subsection{Scattering matrix}\label{sec:S-matrix_continuous} 

We now define the scattering matrix ${\bf S}$.
In a scattering problem, the total field $E_z$ satisfies Eq.~\eqref{eq:A_psi} with no source
and can be written as $E_z(x,y) = E_z^{\rm in}(x,y) + E_z^{\rm sca}(x,y)$.
Here, $E_z^{\rm in}$ is the incident field, and $E_z^{\rm sca}$ is the scattered field that satisfies an outgoing boundary condition at infinity ({\it i.e.}, when $|x| \to \infty$).
To be concrete, we consider light incident from the left at angle $\theta_a$, with
\begin{equation}
\label{eq:Ez_inc}
E_z^{\rm in}(x,y) = \frac{E_0}{\sqrt{k_x^{(a, {\rm L})}}}  u_a^{({\rm L})}(y) e^{ i k_x^{(a, {\rm L})} x }
\end{equation} 
for some constant $E_0$. Outside of the scattering region, the total field $E_z$ can be written as a superposition of the propagating and evanescent channels consistent with the outgoing boundary condition, as 
\begin{equation}
\label{eq:psi_tot_outside_continuous}
\frac{E_z(x,y)}{E_0} = 
\begin{cases}
\displaystyle
\frac{u_a^{({\rm L})}(y) }{\sqrt{k_x^{(a, {\rm L})}}} e^{ i k_x^{(a, {\rm L})} x }
+ \sum_{b} r_{ba}^{({\rm L})} \frac{u_b^{({\rm L})}(y) }{\sqrt{k_x^{(b, {\rm L})}}} e^{ -i k_x^{(b, {\rm L})} x } 
+ \sum_{b}{\vphantom{\sum}}' \tilde{r}_{ba}^{({\rm L})} \frac{u_b^{({\rm L})}(y)}{\sqrt{k_x^{(b, {\rm L})}}} e^{ -i k_x^{(b, {\rm L})} x } , & x \le 0, \\
\displaystyle
\sum_{b} t_{ba}^{({\rm L})}\frac{u_b^{({\rm R})}(y) }{\sqrt{k_x^{(b, {\rm R})}}} e^{ i k_x^{(b, {\rm R})} (x-L) } 
+ \sum_{b}{\vphantom{\sum}}' \tilde{t}_{ba}^{({\rm L})} \frac{u_b^{({\rm R})}(y)}{\sqrt{k_x^{(b, {\rm R})}}} e^{ i k_x^{(b, {\rm R})} (x-L) } , & x \ge L.
\end{cases}
\end{equation}
Summations $\sum_b$ sum over the $N_{\rm L} \approx 2\sqrt{\varepsilon_{\rm L}}W/\lambda$ and $N_{\rm R} \approx 2\sqrt{\varepsilon_{\rm R}}W/\lambda$ propagating channels on the left and right, and summations $\sum'_b$ sum over the countably infinite number of evanescent channels.
Here, $r_{ba}^{({\rm L})}$ and $t_{ba}^{({\rm L})}$ are the reflection and transmission coefficients with input in channel $a$ from the left and output in channel $b$ on the left or right,
normalized by the longitudinal flux, with reference planes on $x=0$ and $x=L$ respectively.
In the absence of absorption or gain ({\it i.e.}, when $\varepsilon_{\rm r}(x,y)$ is real-valued everywhere), flux conservation requires that 
$\sum_{b} |r_{ba}|^2 +
\sum_{b} |t_{ba}|^2 = 1$ for any propagating channel $a$.
Meanwhile, $\tilde{r}_{ba}$ and $\tilde{t}_{ba}$ characterize the evanescent response in the near field.

While it is possible to include the evanescent response in the scattering matrix~\cite{2000_Carminati_PRA}, in most scenarios only the propagating ones are of interest.
We define reflection matrix ${\bf{r}_{\rm{L}}}$ and transmission matrix ${\bf{t}_{\rm{L}}}$ with incidence from the left by their matrix elements $r_{ba}^{({\rm L})}$ and $t_{ba}^{({\rm L})}$; together they form the scattering matrix ${\bf{S}}_{\rm L} = [{{\bf{r}}_{\rm{L}}};{{\bf{t}}_{\rm{L}}}]$ to include output to both sides. In general, we can consider incident light from either left or right, with the full scattering matrix being
\begin{equation}
\label{eq:S_matrix}
{\bf{S}} = \left[ {\begin{array}{*{20}{c}}
{{{\bf{r}}_{\rm{L}}}}&{{{\bf{t}}_{\rm{R}}}}\\
{{{\bf{t}}_{\rm{L}}}}&{{{\bf{r}}_{\rm{R}}}}
\end{array}} \right].
\end{equation}
This full scattering matrix has size $(N_{\rm L}+N_{\rm R})$-by-$(N_{\rm L}+N_{\rm R})$ and is unitary in the absence of absorption or gain.

\subsection{Fisher--Lee relation \label{sec:Fisher_Lee}} 

%There are different ways to solve the scattering problem; one can introduce sources to the right-hand side of Eq.~\eqref{eq:A_psi}, using volume-sources, dipole sources (as in the total-field/scattered-field formalism), or line sources.
%Here we adopt line sources for the solution.

Here we introduce line sources to the right-hand side of Eq.~\eqref{eq:A_psi} to solve the scattering problem.
We first define the retarded Green's function $G(x,y; x',y')$ of the scattering medium, which is the solution of Eq.~\eqref{eq:A_psi} with a point source at $(x',y')$, 
\begin{equation}
\label{eq:A_G}
\lim_{\eta \to 0^+} \left[ - \nabla_{xy}^2 - \frac{\omega^2}{c^2} \varepsilon_{\rm{r}}(x,y) - i \eta \right] G(x,y; x',y') = \delta(x-x')\delta(y-y').
\end{equation}
The infinitesimal absorption $\eta$ imposes an outgoing boundary condition.
We can then use the retarded Green's function to express the total field $E_z$ arising from the incident field in Eq.~\eqref{eq:Ez_inc}, as
\begin{subequations}
\label{eq:psi_tot_continuous}
\begin{align}
\label{eq:psi_tot_1_continuous}
&E_z(x,y) = E_z^{\rm io}(x,y) + E_z^{\rm out}(x,y), \\
\label{eq:psi_io_continuous}
&{E_z^{\rm io}(x,y)}/{E_0} = \frac{1}{\sqrt{k_x^{(a, {\rm L})}}} u_a^{({\rm L})}(y) \left( e^{ i k_x^{(a, {\rm L})} x } - e^{ -i k_x^{(a, {\rm L})} x } \right) H(-x), \\
\label{eq:psi_out_continuous}
&{E_z^{\rm out}(x,y)}/{E_0} = -2i \sqrt{k_x^{(a, {\rm L})}} \int_0^W dy' G(x,y; x'=0,y') u_a^{({\rm L})}(y') ,
\end{align} 
\end{subequations}
where $H(x) = [{\rm sign}(x) + 1 ]/2$ is a Heaviside step function.
The $E_z^{\rm out}$ in Eq.~\eqref{eq:psi_out_continuous} is the solution of Eq.~\eqref{eq:A_psi} with a line source $-2i \sqrt{k_x^{(a, {\rm L})}} \delta(x') u_a^{({\rm L})}(y') E_0$ at $x'=0$ with an outgoing boundary condition.
In the absence of scatterers [{\it i.e.}, when 
$\varepsilon_{\rm{r}}(x,y) = \varepsilon_{\rm L}$
everywhere], such a line source generates outgoing fields
$E_z^{\rm out}(x,y) = \frac{E_0}{\sqrt{k_x^{(a, {\rm L})}}}  u_a^{({\rm L})}(y) e^{ i k_x^{(a, {\rm L})} |x|}$ that propagate away from $x=0$ toward both sides; the addition of $E_z^{\rm io}$ in Eq.~\eqref{eq:psi_io_continuous} subtracts the outgoing wave on the $x<0$ side and adds the incident wave there, such that $E_z^{\rm io} + E_z^{\rm out}$ equals the incident field $E_z^{\rm in}$ in Eq.~\eqref{eq:Ez_inc}.
In the presence of scatterers, $E_z^{\rm out}$ additionally includes the scattered field $E_z^{\rm sca}$ produced by such $E_z^{\rm in}$ interacting with the scatterers.
The $E_z^{\rm io}$ in Eq.~\eqref{eq:psi_io_continuous} is a solution of Eq.~\eqref{eq:A_psi} with an opposite line source $2i \sqrt{k_x^{(a, {\rm L})}} \delta(x') u_a^{({\rm L})}(y') E_0$, so the sum 
$E_z^{\rm io} + E_z^{\rm out}$ satisfies the source-less Eq.~\eqref{eq:A_psi} everywhere; it also satisfies the boundary condition of the scattering problem
%everywhere, since the first term of $E_z^{\rm io}$ is the incident field in Eq.~\eqref{eq:Ez_inc}, and all of the other terms satisfy the outgoing boundary condition at infinity.
defined by $E_z = E_z^{\rm in} + E_z^{\rm sca}$ 
%Therefore, Eq.~\eqref{eq:psi_tot_continuous} is the solution of the scattering problem specified by Eqs.~\eqref{eq:A_psi}
with $E_z^{\rm in}$ from Eq.~\eqref{eq:Ez_inc} and an outgoing $E_z^{\rm sca}$.

By equating Eq.~\eqref{eq:psi_tot_outside_continuous} and Eq.~\eqref{eq:psi_tot_continuous}, evaluating it at $x=0$ and $x=L$ (note that only $E_z^{\rm out}$ contributes since $E_z^{\rm io}=0$ at $x=0$ and $x=L$), and projecting onto the output channels using the orthonormality of the transverse modes, we obtain the reflection and transmission coefficients as 
\begin{subequations}
\label{eq:Fisher-Lee_continuous}
\begin{align}
\label{eq:r_continuous}
r_{ba}^{({\rm L})} &= -2 i \sqrt{ k_x^{(b, {\rm L})} k_x^{(a, {\rm L})} } %G_{ba}(0,0),
\int d y \int d y' u_b^{(\rm L)*}(y) G(x=0, y; x'=0, y')  u_a^{(\rm L)}(y') -\delta_{ba}, \\
\label{eq:t_continuous}
t_{ba}^{({\rm L})} &= -2 i \sqrt{ k_x^{(b, {\rm R})} k_x^{(a, {\rm L})} } %G_{ba}(L,0), \\
\int d y \int d y' u_b^{(\rm R)*}(y) G(x=L, y; x'=0, y')  u_a^{(\rm L)}(y'),
\end{align}
\end{subequations}
which is known as the Fisher--Lee relation, initially derived for the single-particle Schr\"odinger equation in the context of quantum transport~\cite{1981_Fisher_PRB, Datta1995, Wimmer2009}.
Eq.~\eqref{eq:Fisher-Lee_continuous} can be understood intuitively: 
for each incident angle $\theta_a$, $-2i \sqrt{k_x^{(a, {\rm L})}} \delta(x') u_a^{({\rm L})}(y')$ is the input source, $\sqrt{k_x^{(b, {\rm L})}} \delta(x) u_b^{({\rm L})}(y)$ and $\sqrt{k_x^{(b, {\rm R})}} \delta(x-L) u_b^{({\rm R})}(y)$ are the output projections for reflection and transmission coefficients into different outgoing angles $\theta_b$, and the $\delta_{ba}$ in Eq.~\eqref{eq:r_continuous} is the projection of the incident field $E_z^{\rm in}$ on the incident side.
Eq.~\eqref{eq:Fisher-Lee_continuous} is a special case of Eq.~(2) in the main text, written for the specific geometry considered here prior to discretization.

\section{Finite-difference discretization \label{sec:dis_Fisher_Lee}}

We need to discretize the system in order to compute its scattering matrix numerically.
Here we consider finite-difference discretization on the Yee grid~\cite{Yee1966_TAP}.
On the Yee grid, the first derivatives are approximated with center difference with second-order accuracy, and the grid points of different field components are staggered in space.
For 2D TM waves, we put the discretized $E_z$ component at
\begin{equation}
\label{eq:E_z_FD}
E_z(x=x_n, y=y_m) \to E_{z (n,m)}
\end{equation}
where
\begin{equation}
\label{eq:xy_FD}
x_n\equiv\left(n-\frac{1}{2}\right)\Delta x, \quad y_m\equiv\left(m-\frac{1}{2}\right) \Delta x
\end{equation}
with $(n,m)$ being integer indices and $\Delta x$ the discretization grid size.
The $H_x$ component is located at $(x_n, y_{m+1/2})$, and the $H_y$ component is located at $(x_{n+1/2}, y_m)$.
This gives the discretized second derivative of $E_z$ in $x$ as
\begin{equation}
\label{eq:ddx2_FD}
\left. \frac{\partial^2 }{\partial x^2} E_z(x,y) \right|_{x=x_n, y=y_m}
\to \frac{E_{z (n+1,m)} - 2E_{z(n,m)} + E_{z (n-1,m)}}{\Delta x^2},
\end{equation}
and similarly for $\partial^2 E_z/\partial y^2$.
We discretize the relative permittivity profile $\varepsilon_{\rm r}(x,y)$ through subpixel smoothing~\cite{2006_Farjadpour_OL}; for TM waves, this corresponds simply to averaging $\varepsilon_{\rm r}(x,y)$ within the $\Delta x^2$ area of each cell centered at $(x_n,y_m)$, as
\begin{equation}
\varepsilon_{n,m} =
\frac{1}{\Delta x^2}
\int_{(n-1)\Delta x}^{n\Delta x} dx
\int_{(m-1)\Delta x}^{m\Delta x} dy
\, \varepsilon_{\rm r}(x,y).
\end{equation}
Note that different from the notation of Ref.~\cite{Yee1966_TAP}, in Eq.~\eqref{eq:xy_FD} we introduced the half-pixel offset so that the first pixel $(n,m)=(1,1)$ of $E_z$ will have its lower corner at $(x,y)=(0,0)$; this is more convenient when we only deal with $E_z$.
Then, the differential operator $- \nabla_{xy}^2 - {\left( {\omega /c} \right)^2} \varepsilon_{\rm r}(x,y)$ of Eq.~\eqref{eq:A_psi} is discretized into $A_{(n,m),(n',m')}/\Delta x^2$ with
\begin{equation}
\label{eq:A_FD}
A_{(n,m),(n',m')} = - (\delta_{n+1,n'} + \delta_{n-1,n'}) \delta_{m,m'} + (4 - \beta^2 \varepsilon_{n,m}) \delta_{n,n'} \delta_{m,m'} - ( \delta_{m+1,m'} + \delta_{m-1,m'} ) \delta_{n,n'}
\end{equation}
away from the boundaries, with $\beta \equiv (\omega/c)\Delta x$.
This is the finite-difference frequency-domain (FDFD) formulation.

In practice, the system size is made finite, and a single index is used to go through both indices $(n,m)$.
With that single index, $E_{z (n,m)}$ becomes an $N' \times 1$ column vector $E_z$, and $A_{(n,m),(n',m')}$ becomes an $N' \times N'$ square matrix ${\bf A}$.
Then, Eq.~\eqref{eq:A_psi} with no source becomes a finite-sized matrix-vector multiplication, 
\begin{equation}
\label{eq:AE_z_FD}
{\bf A} E_z = 0.
\end{equation}

Now we consider the two-sided geometry in Sec.~\ref{sec:s_matrix}.
We restrict the integer index $m$ to $1 \le m \le n_y$ with $n_y \equiv W/\Delta x$ being the transverse number of grid points (discretizing the region $0 \le y \le W$); $\Delta x$ can be chosen such that $n_y$ is an integer.
The Bloch periodic boundary condition 
$E_{z, (n,n_y+1)} = E_{z, (n,1)} e^{i k_{\rm B} n_y \Delta x}$
and
$E_{z, (n,0)} = E_{z, (n,n_y)} e^{-i k_{\rm B} n_y \Delta x}$
manifests in the boundary elements of matrix ${\bf A}$.
The system size in $x$ is infinite, but we must truncate it to a finite size with an effective open boundary with no reflection at the interface of truncation; we do so with the perfectly matched layer (PML)~\cite{2005_Gedney_book_chapter}, which attenuates the outgoing waves with minimal reflection.
In this way, the outgoing boundary condition is also built into matrix ${\bf A}$.
Also, matrix ${\bf A}$ and vector $E_z$ both have finite sizes.

The retarded Green’s function $G(x,y; x',y')$ defined in Eq.~\eqref{eq:A_G} is the inverse of the differential operator.
When discretized, it becomes a matrix and is simply given by the matrix inverse
\begin{equation}
\label{eq:A_inv}
 {\bf G} = {\bf A}^{-1}.
\end{equation}
The outgoing boundary condition is automatic since it is already built into matrix ${\bf A}$.

Under subpixel smoothing, Eq.~\eqref{eq:syst} becomes
\begin{equation}
\label{eq:epsilon_slab_FD}
\varepsilon_{n,m} =
\begin{cases}
\, \varepsilon_{\rm L}, & n \le 0, \quad \quad \quad \ 1 \le m \le n_y, \\
\, \varepsilon_{n,m} , & 1 \le n \le n_x,\quad 1 \le m \le n_y, \\
\, \varepsilon_{\rm R}, & n \ge n_x + 1, \quad 1 \le m \le n_y,
\end{cases}
\end{equation}
where $n_x \equiv \lceil L/\Delta x \rceil$ is the number of grid points for the scattering region.
The rest follows the same steps as in Sec.~\ref{sec:s_matrix}, which we summarize below.

\subsection{Channels in discrete homogeneous space\label{sec:homo_inf}}

Given the discrete translational symmetry in $n$ when $n \le 0$ and $n \ge n_x + 1$, Eq.~\eqref{eq:Ez_a} still holds in the form of
\begin{equation}
\label{eq:phi_a_FD}
E^{(a,\pm)}_{z(n,m)} = \frac{1}{{\sqrt {{\nu _a}} }} u_{ma} \exp\left[ \pm i k_x^{(a)} \Delta x \left(n- \frac{1}{2}\right) \right],
\end{equation}
with the longitudinal-flux normalization factor ${{\nu _a}}$ to be determined.
The transverse modes of Eq.~\eqref{eq:phi_a} become
\begin{equation}
\label{eq:phi_FD}
u_{ma} = \frac{1}{{\sqrt {{n_y}} }}\exp \left[ {ik_y^{\left( a \right)}\Delta x\left( {m - {m_0}} \right)} \right],
\end{equation}
which make up matrix ${\bf u}$, with
$y_0 = (m_0 - 1/2)\Delta x$. The transverse wave number is still $k_y^{(a)} = k_{\rm B} + a \frac{2\pi}{n_y \Delta x}$, but channel $a$ is now equivalent to channel $a + n_y$ due to aliasing, so there are only $n_y$ distinct channels in total (propagating ones plus evanescent ones) instead of an infinite number of them. Completeness and orthonormality of the transverse modes means that the $n_y \times n_y$ matrix ${\bf u}$ is unitary when all $n_y$ channels are included.
Inserting Eqs.~\eqref{eq:phi_a_FD}--\eqref{eq:phi_FD} into Eq.~\eqref{eq:AE_z_FD} yields the finite-difference dispersion relation
\begin{equation}
\label{eq:dispersion}
{\varepsilon _{{\rm{bg}}}}{\left( {\frac{\omega }{c}} \right)^2}\Delta {x^2} = 4 \, {\sin ^2}\left( {\frac{{k_x^{(a)}\Delta x}}{2}} \right) + 4 \, {\sin ^2}\left( {\frac{{k_y^{(a)}\Delta x}}{2}} \right).
\end{equation}

Flux orthogonality continues to hold in the discretized system.
To preserve flux conservation ({\it i.e.}, Poynting's theorem), the Poynting vector in the discrete system should be defined using the product of an ${\bf E}$ field component and an ${\bf H}$ field component at $\Delta x/2$ away~\cite{1994_Chew_JAP}.
For 2D TM felds, this means that the $x$-directional flux is proportional to
${\rm{Im}}\left[ {E_{z\left( {n,m} \right)}^*{E_{z\left( {n + 1,m} \right)}}} \right]$.
For each channel in Eq.~\eqref{eq:phi_a_FD}, such flux
is proportional to $\frac{\pm1}{{{{\left| { {{\nu _a}} } \right|}}}} \sin \left( {k_x^{\left( a \right)}\Delta x} \right)$ for propagating channels (where $k_x^{(a)}$ is real-valued), zero for evanescent channels (where $k_x^{(a)}$ is imaginary or complex-valued). Therefore, we choose the the flux normalization factor in Eq.~\eqref{eq:phi_a_FD} to be
\begin{equation}
\label{eq:mu_a}
\nu_a = \sin\left(k_x^{(a)} \Delta x \right)
\end{equation}
so that all propagating channels carry equal longitudinal flux.

\subsection{Finite-difference Fisher--Lee relation}

The definition of the scattering matrix is the same as the continuous case in Eq.~\eqref{eq:psi_tot_outside_continuous}, simply with $x$ and $y$ replaced through Eq.~\eqref{eq:xy_FD}.

The derivation of the scattering matrix in terms of the Green's function (namely, the Fisher--Lee relation) is the same as the continuous case except for one difference: the sources and the projections were placed at $x=0$ and $x=L$ in the continuous case, but the corresponding spatial indices $n = (x/\Delta x) + (1/2)$ would lie on $n=1/2$ and $n = (L/\Delta x) + (1/2)$, which are generally not integer points.
Therefore, for the discretized system, we put the sources and the projections at integer indices $n=0$ and $n=n_x + 1$ (recall that $n_x \equiv \lceil L/\Delta x \rceil$). This gives the discrete version of Eq.~\eqref{eq:Fisher-Lee_continuous} as
\begin{subequations}
\label{eq:Fisher-Lee_FD}
\begin{align}
\label{eq:r_FD}
r_{ba}^{({\rm L})} &= \left[ -2 i \sqrt{ \nu_b^{({\rm L})} \nu_a^{({\rm L})}}
\sum_{m}   \sum_{m'}  
u_{mb}^{({\rm L})*}
G_{(n=0,m),(n'=0,m')}
u_{m'a}^{({\rm L})} -\delta_{ba} \right] \delta_b^{({\rm L})} \delta_a^{({\rm L})}, \\
\label{eq:t_FD}
t_{ba}^{({\rm L})} &= \left[-2 i \sqrt{ \nu_b^{({\rm R})} \nu_a^{({\rm L})}}
\sum_{m}   \sum_{m'}  
u_{mb}^{({\rm R})*}
G_{(n=n_x+1,m),(n'=0,m')}
u_{m'a}^{({\rm L})} \right] \delta_b^{({\rm R})} \delta_a^{({\rm L})}.
\end{align}
\end{subequations}
Here, $\delta_b^{({\rm L/R})} = \exp\left[ - i k_x^{(b, {\rm L/R})} \Delta x \delta n^{({\rm L/R})} \right]$ and $ \delta_a^{({\rm L/R})} = \exp\left[ - i k_x^{(a, {\rm L/R})} \Delta x \delta n^{({\rm L/R})} \right]$ are phase factors that compensate for the shift of the detection plane and the source plane respectively, with $\delta n^{({\rm L})} = 1/2$,
$\delta n^{({\rm R})} = 1/2 + \lceil L/\Delta x \rceil - (L/\Delta x)$;
these factors can be ignored if the location of the reference plane is not of interest.

Based on Eq.~\eqref{eq:A_inv}, Eq.~\eqref{eq:Fisher-Lee_FD}, and considering inputs from both sides as in Eq.~\eqref{eq:S_matrix}, we can write the full scattering matrix ${\bf S}$ as
\begin{equation}
\label{eq:S_CAB_D}
{\bf{S}} = {\bf{C}}{{\bf{A}}^{ - 1}}{\bf{B}} - {\bf{D}},
\end{equation}
which is unitary in the absence of absorption or gain if all propagating channels are included.
A schematic illustration of Eq.~\eqref{eq:S_CAB_D} is given in Fig.~1c of the main text.

Matrix ${\bf A}$ is the discrete version of the differential operator $- \nabla_{xy}^2 - {\left( {\omega /c} \right)^2} \varepsilon_{\rm r}(x,y)$ in Eq.~\eqref{eq:A_psi}, given by Eq.~\eqref{eq:A_FD}, PML, and the boundary conditions.
The input matrix ${\bf{B}}$ is
\begin{equation}
\label{eq:B_matrix}
{\bf{B}} = \left[ {\begin{array}{*{20}{c}}
{\bf{0}}&{\bf{0}}\\
{{{\bf{B}}_{\rm{L}}}}&{\bf{0}}\\
{\bf{0}}&{\bf{0}}\\
{\bf{0}}&{{{\bf{B}}_{\rm{R}}}}\\
{\bf{0}}&{\bf{0}}
\end{array}} \right].
\end{equation}
The top (bottom) block row of zeros correspond to indices $n < 0$ ($n > n_x + 1$) with PML and homogeneous space on the left (right); they are shown in green in Fig.~1c--d of the main text.
The second (fourth) block row corresponds to index $n=0$ ($n=n_x+1$), which is the left (right) surface where input sources are placed; they are shown in red in Fig.~1c--d.
The third block row corresponds to the scattering region with indices $1 \le n \le n_x$, shown in blue in Fig.~1c--d.
Matrices ${{\bf{B}}_{\rm{L}}}$ and ${{\bf{B}}_{\rm{R}}}$ have sizes $n_y \times M_{{\rm {L}}}$ and $n_y \times M_{{\rm {R}}}$; they are line sources on the surface, given by 
\begin{equation}
\label{eq:B_L}
{{\bf{B}}_{\rm{L}}} = -2i{{\bf{u }}_{\rm{L}}}\sqrt {{{\bf{\nu }}_{\rm{L}}}} \delta^{({\rm L})}, \quad
{{\bf{B}}_{\rm{R}}} = -2i{{\bf{u }}_{\rm{R}}}\sqrt {{{\bf{\nu }}_{\rm{R}}}} \delta^{({\rm R})},
\end{equation}
where the $n_y \times M_{{\rm {L}}}$ matrix ${\bf u}_{{\rm{L}}}$ is defined in Eq.~\eqref{eq:phi_FD},
matrices $\sqrt {{\nu _{{\rm{L}}}}}  = {\rm{diag}}\left( \left\{ {\sqrt {\nu _a^{\left( {{\rm{L}}} \right)}} } \right\}_a \right)$ and $ \delta^{({\rm L})}  = {\rm{diag}}\left( \left\{ \delta_a^{({\rm L})} \right\}_a \right)$ are $M_{{\rm {L}}} \times M_{{\rm {L}}}$ diagonal matrices for flux normalization and phase shift respectively, and similarly with ${\bf B}_{\rm R}$.
Similarly, the output matrix ${\bf{C}}$ performs the projection onto the output channels,
\begin{equation}
\label{eq:C_matrix}
{\bf{C}} = \left[ {\begin{array}{*{20}{c}}
{\bf{0}}&{{{\bf{C}}_{\rm{L}}}}&{\bf{0}}&{\bf{0}}&{\bf{0}}\\
{\bf{0}}&{\bf{0}}&{\bf{0}}&{{{\bf{C}}_{\rm{R}}}}&{\bf{0}}
\end{array}} \right],
\end{equation}
with 
\begin{equation}
\label{eq:C_L}
{{\bf{C}}_{\rm{L}}} = \delta^{({\rm L})} \sqrt {{{\bf{\nu }}_{\rm{L}}}}{{\bf u}_{\rm{L}}^\dagger}, \quad 
{{\bf{C}}_{\rm{R}}} = \delta^{({\rm R})} \sqrt {{{\bf{\nu }}_{\rm{R}}}}{{\bf u}_{\rm{R}}^\dagger},
\end{equation}
where $^\dagger$ stands for conjugate transpose.
Note that the list of the $M'=M'_{\rm L} + M'_{\rm R}$ output channels do not have to be the same as the list of the $M=M_{\rm L} + M_{\rm R}$ input channels, but for simplicity we do not introduce a separate notation.
Matrix ${\bf{D}}$ here is the $\delta_{ba}$ in the Fisher--Lee relation, with elements that equal $\delta_b^{({\rm L/R})} \delta_a^{({\rm L/R})}$ 
when the input channel is the same as the output channel, 0 otherwise.

For the numerical computations and implementation, it is faster and simpler to take the prefactor $-2i$ in Eq.~\eqref{eq:B_L} and the $\sqrt {{{\bf{\nu }}_{\rm{L/R}}}} \delta^{({\rm L/R})}$ prefactors in Eq.~\eqref{eq:B_L} and Eq.~\eqref{eq:C_L} out of ${\bf{B}}_{{\rm{L}}/{\rm{R}}}$ and ${{\bf{C}}_{{\rm{L}}/{\rm{R}}}}$, 
and multiply such prefactors after ${\bf C}{\bf A}^{-1}{\bf B}$ is computed.
This means we can use
\begin{equation}
\label{eq:BC_no_prefactor}
{{\bf{B}}_{\rm{L}}} = {\bf u}_{\rm{L}}, \quad
{{\bf{B}}_{\rm{R}}} = {\bf u}_{\rm{R}}, \quad
{{\bf{C}}_{\rm{L}}} = {\bf u}_{\rm{L}}^\dagger, \quad 
{{\bf{C}}_{\rm{R}}} = {\bf u}_{\rm{R}}^\dagger, \quad 
\end{equation}
instead for the numerical computations.
We omit {\bf{D}} from matrix ${\bf K}$ [as in Eq.~\eqref{eq:K_matrix} below] and subtract {\bf{D}} from ${\bf C}{\bf A}^{-1}{\bf B}$ after the prefactors are put back. 

While the details are system dependent, the concepts above are general, and scattering matrices can always be written in the form of Eq.~\eqref{eq:S_CAB_D} regardless of the discretization scheme, the geometry, the type of sources, and the type of outputs of interest.
In general, ${\bf{D}} = {\bf{C}}{\bf A}_0^{-1}{\bf{B}} - {\bf S}_0$, where ${\bf A}_0$ is the Maxwell operator of a reference system ({\it e.g.}, a homogeneous one) for which the scattering matrix ${\bf S}_0$ is known; in the specific case above, ${\bf{C}}{\bf A}_0^{-1}{\bf{B}} = {\bf I} + {\bf S}_0$ where the identity matrix ${\bf I}$ comes from projection of $E_z^{\rm in}$ on the incident side (assuming the full scattering matrix and ignoring the phase-shift factors to simplify notation).

\section{Utilizing symmetry in augmented partial factorization (APF)\label{sec:symmetrize}}

In APF, a partial factorization is performed on the augmented sparse matrix
\begin{equation}
\label{eq:K_matrix}
{\bf{K}} \equiv \left[ {\begin{array}{*{20}{c}}
{\bf{A}}&{\bf{B}}\\
{\bf{C}}&{\bf{0}}
\end{array}} \right].
\end{equation}
For most linear solvers such as the MUMPS package~\cite{Amestoy2001_SIAM} we use, the computing time and memory usage of such partial factorization can be reduced when matrix ${\bf K}$ is symmetric.
Thanks to reciprocity, the bulk of matrix {\bf{A}} as in Eq.~\eqref{eq:A_FD} is symmetric; 
periodic boundary condition and the use of uniaxial PML does not break such symmetry.
Therefore, matrix ${\bf K}$ will be symmetric if we can make ${\bf{C}} = {\bf{B}}^{\rm{T}}$, or equivalently ${{\bf{C}}_{{\rm{L}}}} = {\bf{B}}_{{\rm{L}}}^{\rm{T}}$ and ${{\bf{C}}_{{\rm{R}}}} = {\bf{B}}_{{\rm{R}}}^{\rm{T}}$.

From Eq.~\eqref{eq:BC_no_prefactor}, we see that when the list of input channels equals the list of output channels, we would have the desired ${{\bf{C}}_{{\rm{L}}}} = {\bf{B}}_{{\rm{L}}}^{\rm{T}}$ if the transverse mode profiles were real-valued.
The transverse mode profiles in Eq.~\eqref{eq:phi_FD} are not real-valued, but we can see that taking the complex conjugate of the profile is equivalent to flipping the sign of $k_y^{(a)}$;
with a periodic boundary condition in $y$ (where $k_{\rm B} = 0$), this corresponds to flipping the sign of the channel index $a$.
Therefore, we can achieve ${{\bf{C}}_{{\rm{L}}}} = {\bf{B}}_{{\rm{L}}}^{\rm{T}}$ simply by making the list of output channels having the opposite channel index as the list of input channels.
When the full scattering matrix or reflection matrix is computed, all we need to do is to choose a particular ordering of the channels (which can be reversed after ${\bf C}{\bf A}^{-1}{\bf B}$ is computed).
When only a subset of the scattering matrix is needed, we can further pad input and/or output channels to achieve ${{\bf{C}}_{{\rm{L}}}} = {\bf{B}}_{{\rm{L}}}^{\rm{T}}$ to make matrix ${\bf K}$ symmetric.
Note this is applicable to any structure and does not require any symmetry in ${\varepsilon _{\rm{r}}}(x,y)$.

\begin{figure*}[t]
\includegraphics[width=1.0\textwidth]{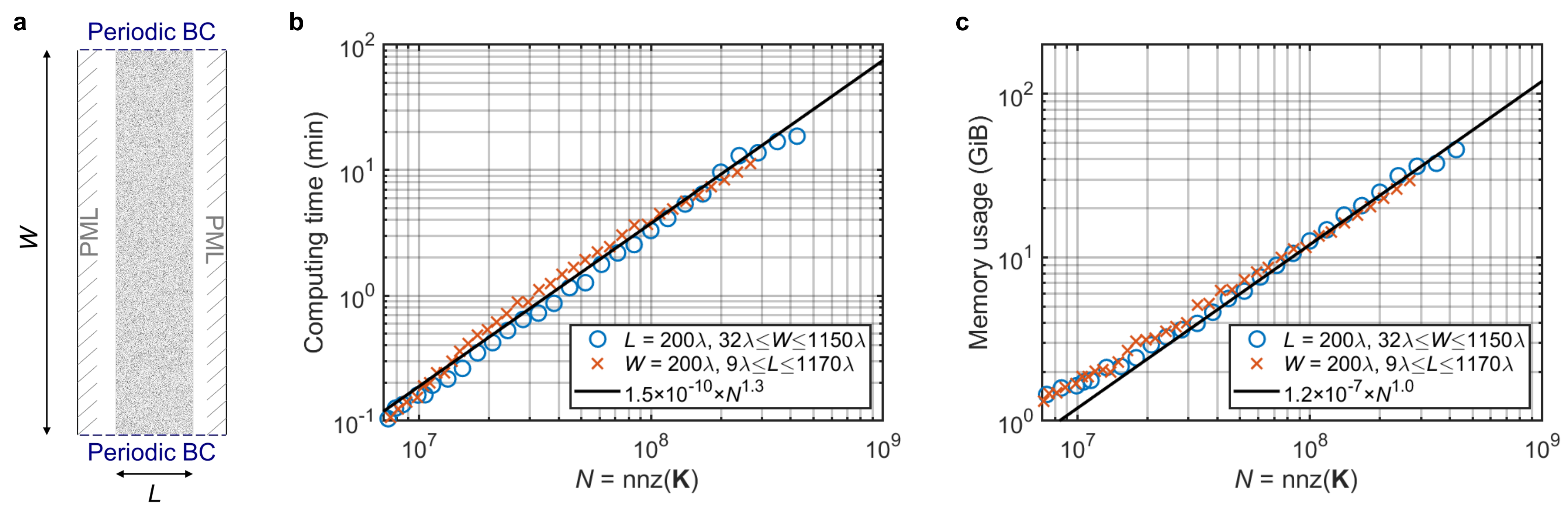}
\caption{\label{fig:scaling_APF} {\bf Computing time and memory usage of APF}. {\bf{a}}, Schematic of the disordered system considered, with thickness $L$ and width $W$. Other parameters are the same as in Fig.~2 of the main text. {\bf{b}--c}, Computing time ({\bf{b}}) and memory usage ({\bf{c}}) versus the number $N$ of nonzero elements in matrix {\bf{K}}. Symbols are from simulations; black lines are fitting curves.}
\end{figure*}

\section{Computing time and memory scaling of APF \label{sec:APF_nnz}}

We implement APF under finite-difference discretization for 2D TM fields as described above, and here we map out its computing time and memory usage scaling as a function of the system size. We consider the same disordered system as in Fig.~2 of the main text but with different system sizes: (1) fixing thickness at $L = 200 \lambda$ while varying width $W$ from 32 $\lambda$ to 1150 $\lambda$, and (2) fixing $W = 200 \lambda$ while varying $L$ from 9 $\lambda$ to 1170 $\lambda$. %, with $\lambda$ being the vacuum wavelength. 
The computing time and memory usage are obtained with serial computations on Intel Xeon Gold 6130 nodes.
For each system size, the full scattering matrix is computed 10 times; the average computing time and the maximal recorded memory usage among the 10 computations is used.
A constant 0.57 GiB memory used by MATLAB R2020b is subtracted from the memory usage.

The computing time and memory usage are well characterized by the number $N$ of nonzero elements in the sparse matrix {\bf{K}} [denoted by ${\rm{nnz}}({\bf{K}})$], as shown in Fig.~\ref{fig:scaling_APF}b--c. An $\mathcal{O}(N^{1.3})$ curve and an $\mathcal{O}(N)$ curve closely describe the computing time and the memory usage for all of these systems.

Here, ${\rm{nnz}}({\bf{K}})$ is dominated by ${\rm{nnz}}({\bf{A}})$, with {\rm{nnz}}({\bf{A}})/{\rm{nnz}}({\bf{B}}) in between 1 and  100 among these systems.
Therefore, ${\rm{nnz}}({\bf{K}}) \approx {\rm{nnz}}({\bf{A}})$, which is proportional to the number of pixels in the discretization and independent of the number $M$ of input channels.

\section{APF with compressed input/output matrices (APF-c) \label{sec:APF_c}}

When ${\rm{nnz}}({\bf{B}})$ and/or ${\rm{nnz}}({\bf{C}})$ exceeds ${\rm{nnz}}({\bf{A}})$, the computing time and memory usage of APF can grow with the number of inputs $M$, which is not desirable.
But there is a simple solution: we can ``compress'' matrices ${\bf B}$ and ${\bf C}$ to reduce their number of nonzero elements, perform the partial factorization, and then ``decompress.''
Conceptually, this is similar to other forms of data compression.
We call this APF-c with c standing for compression.

In this section, we consider a simple compression scheme based on the Fourier transform. This Fourier scheme is more than sufficient for the metasurface examples considered in this paper; note when the accuracy is sufficient and when ${\rm{nnz}}({\bf{B}})$ and ${\rm{nnz}}({\bf{C}})$ goes below ${\rm{nnz}}({\bf{A}})$, further compression is no longer necessary. 
More advanced compression strategies~\cite{Sayood_2017_book} may be used if there is such need.

\begin{figure*}[t]
\includegraphics[width=0.7\textwidth]{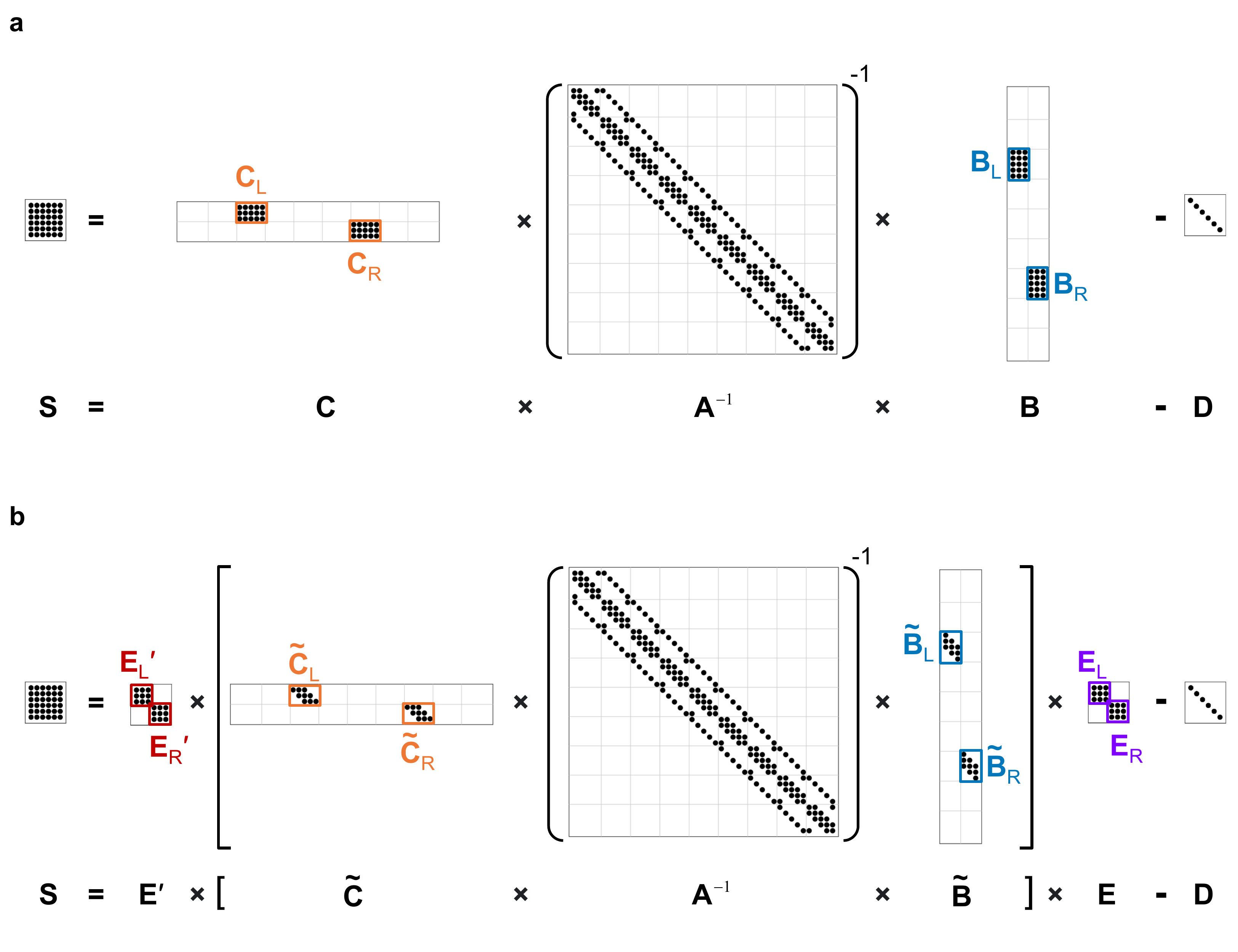}
\caption{\label{fig:APF_c} {\bf Concept of APF with compression (APF-c)}. {\bf{a}}, Schematic illustration of Eq.~\eqref{eq:S_CAB_D} that highlights the dense blocks ${{\bf{B}}_{{\rm{L}}}}$, ${{\bf{B}}_{{\rm{R}}}}$ and ${{\bf{C}}_{{\rm{L}}}}$, ${{\bf{C}}_{{\rm{R}}}}$ of the input and output matrices.
{\bf{b}}, With APF-c, the input and output matrices are transformed such that ${\bf{\tilde B}}_{\rm{L}}$, ${\bf{\tilde B}}_{\rm{R}}$, ${\bf{\tilde C}}_{\rm{L}}$, ${\bf{\tilde C}}_{\rm{R}}$ are spatially localized (see Fig.~\ref{fig:phiQL_comp}) and can be truncated with minimal or no loss of accuracy. The transformations are reversed after $\tilde{\bf C} {\bf A}^{-1} \tilde{\bf B}$ is computed.}
\end{figure*}

\begin{figure*}[t]
\includegraphics[width=1\textwidth]{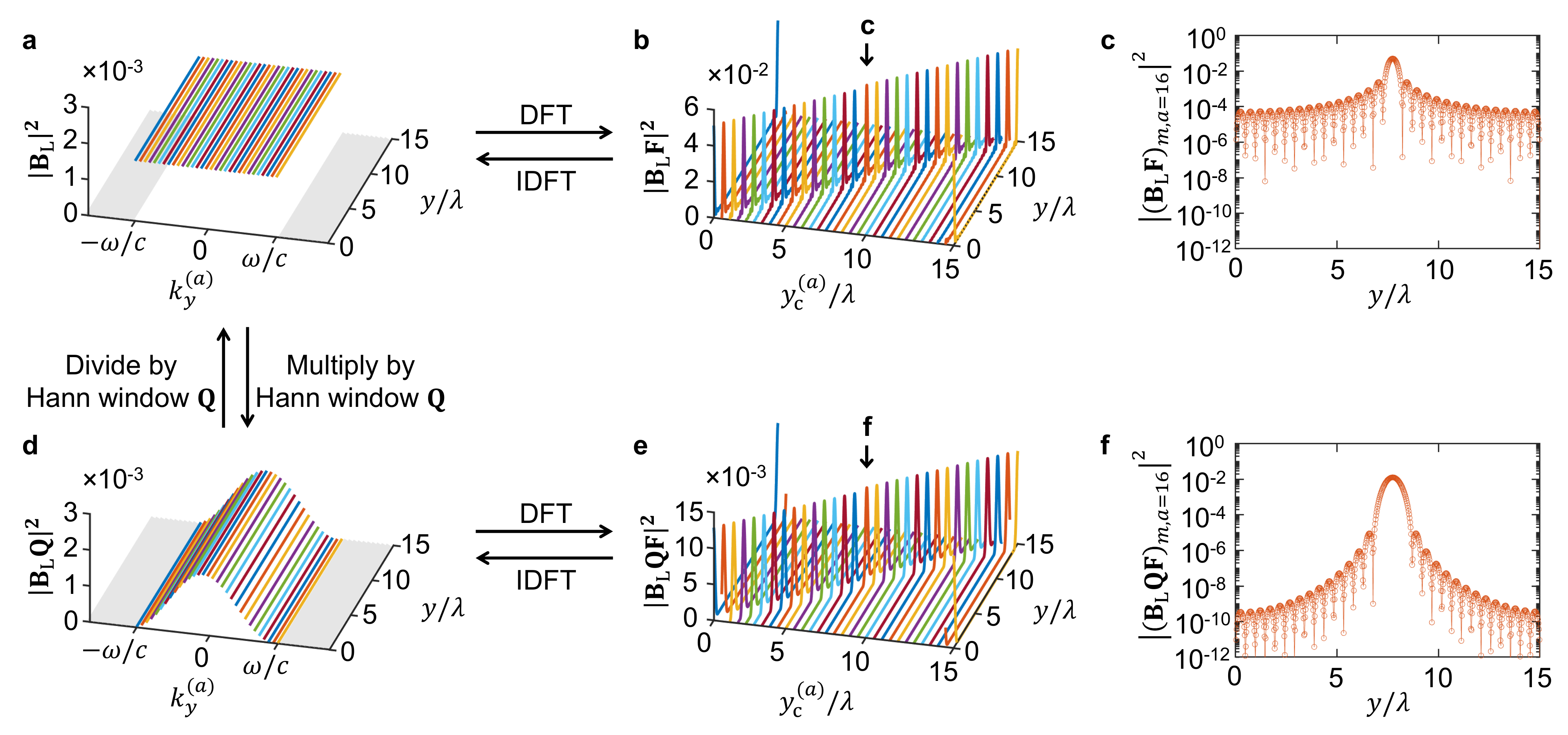}
\caption{\label{fig:phiQL_comp} {\bf{Transformations on the input matrix in APF-c.}}
{\bf{a}}, Original input matrix ${\bf B}_{\rm L}$; each column spans the full width of the system.
{\bf{b}}, Input matrix after a discrete Fourier transform, ${\bf{\tilde B}}_{\rm{L}} = {\bf B}_{\rm L} {\bf F}$; each column is now spatially localized.
{\bf{c}}, The central column of ${\bf{\tilde B}}_{\rm{L}}$ in {\bf{b}}, in log scale.
{\bf{d}}, Input matrix ${\bf B}_{\rm L} {\bf Q}$ weighted by the Hann window function.
{\bf{e}}, Input matrix after Hann window and discrete Fourier transform, ${\bf{\tilde B}}_{\rm{L}} = {\bf B}_{\rm L} {\bf Q} {\bf F}$.
{\bf{f}}, The central column of ${\bf{\tilde B}}_{\rm{L}}$ in {\bf{e}}, in log scale; the Hann window makes the tails decay faster.
The system considered here has width $W = 15\lambda$ discretized with $\Delta x = \lambda/40$. The subscript on the number of channels is dropped to simplify notation. 
}
\end{figure*}

Figure~\ref{fig:APF_c}a illustrates the expression of the scattering matrix {\bf{S}} in Eq.~\eqref{eq:S_CAB_D}, highlighting the nonzero blocks ${{\bf{B}}_{{\rm{L}}}}$, ${{\bf{B}}_{{\rm{R}}}}$ of the input matrix ${\bf B}$ and the nonzero blocks ${{\bf{C}}_{{\rm{L}}}}$, ${{\bf{C}}_{{\rm{R}}}}$ of the output matrix ${\bf C}$. 
These nonzero blocks are placed on the front ($x=0$) and back ($x=L$) surfaces of the scattering region, as indicated in red in Fig.~1d of the main text.
Even though matrices ${\bf B}$ and ${\bf C}$ are nonzero only on these two surfaces, these nonzero blocks are dense and can still contain many elements when the system is wide. Each column of ${\bf B}_{\rm L}$ is the cross section of a plane wave given in Eq.~\eqref{eq:phi_FD}, which covers the full range of $y$ as shown in Fig.~\ref{fig:phiQL_comp}a.  

While a single plane wave is extended in space, a superposition of plane waves can form a sharp focus, with the focus location determined by the relative phases of the constituting plane waves.
And such a conversion between angular basis (plane waves) and spatial basis (focused waves) is invertible.
With this idea in mind, we will take a discrete Fourier transform of the dense block ${\bf B}_{\rm L}$ along its channel index $a$, and similarly with ${{\bf{B}}_{{\rm{R}}}}$, ${{\bf{C}}_{{\rm{L}}}}$, and ${{\bf{C}}_{{\rm{R}}}}$.
For concreteness, below we consider having $M_{\rm L}$ input channels from the left with consecutive channel indices $a =-\frac{{M_{\rm{L}} - 1}}{2}$, $\cdots$, 0, $\cdots$, $\frac{{M_{\rm{L}} - 1}}{2}$, under periodic boundary condition with $m_0=0$.
We define an $M_{\rm{L}}{\times}M_{\rm{L}}$ discrete Fourier transform (DFT) matrix ${{\bf{F}}_{{M_{\rm{L}}}}}$ by its elements,
\begin{equation}
\label{eq:dft}
{\left( {{{\bf F}_{{M_{\rm{L}}}}}} \right)_{ba}} = \frac{1}{{\sqrt {{M_{\rm{L}}}} }}{e^{ - i\frac{{2\pi }}{{{M_{\rm{L}}}}}ba}}.
\end{equation}
The inverse DFT matrix is then ${\bf{F}}_{{M_{\rm{L}}}}^{-1}  = {\bf{F}}_{{M_{\rm{L}}}}^{\dagger}$. 
Note that the indices $b$ and $a$ here are centered around zero; the DFT matrix is commonly defined with indices starting at zero instead, which is equivalent to the definition here after an index shift.

When taking a superposition of plane waves to form a focus, the weight of the constituting plane waves can determine how fast the intensity decays away from the focal spot. Therefore, we also introduce a diagonal matrix ${{\bf{Q}}_{{M_{\rm{L}}}}}$, defined as
\begin{equation}
\label{eq:Q_matrix}
{{\bf{Q}}_{{M_{\rm{L}}}}} = {\rm{diag}}\left( {\left\{ {{q^{M_{\rm{L}}}_{a}}} \right\}} \right),
\end{equation}
where $\left\{ {{q^{M_{\rm{L}}}_{a}}} \right\}_a$ are nonzero real numbers (the weights). 

Then, matrix ${\bf B}_{\rm L}$ can be written as
\begin{equation}
{{\bf{B}}_{\rm{L}}} = {\bf u}_{\rm{L}} =
\underbrace{\left( {{{\bf{u }}_{\rm{L}}}{{\bf{Q}}_{{M_{\rm{L}}}}}{{\bf{F}}_{{M_{\rm{L}}}}}} \right)}_{\displaystyle\equiv{\bf{\tilde B}}_{\rm{L}}}
\underbrace{\left( {{\bf{F}}_{{M_{\rm{L}}}}^{-1} {\bf{Q}}_{{M_{\rm{L}}}}^{ - 1} } \right)}_{\displaystyle\equiv{\bf{E}}_{\rm{L}}}.
\end{equation}
Similarly,
\begin{subequations}
\begin{align}
\label{eq:B_R_tilda_E_R}
{{\bf{B}}_{\rm{R}}} = {\bf u}_{\rm{R}} &= \left( {{{\bf{u }}_{\rm{R}}}{{\bf{Q}}_{{M_{\rm{R}}}}}{{\bf{F}}_{{M_{\rm{R}}}}}} \right)\left( {{\bf{F}}_{{M_{\rm{R}}}}^{-1} {\bf{Q}}_{{M_{\rm{R}}}}^{ - 1} } \right) \equiv {{\bf{\tilde B}}_{\rm{R}}}{{\bf{E}}_{\rm{R}}},\\
\label{eq:E_L_prime_C_L_tilda}
{{\bf{C}}_{\rm{L}}} = {\bf u}_{\rm{L}}^\dag &= \left( { {\bf{Q}}_{{M_{\rm{L}}}}^{ - 1}{{\bf{F}}_{{M_{\rm{L}}}}}} \right)\left( {{\bf{F}}_{{M_{\rm{L}}}}^{-1} {{\bf{Q}}_{{M_{\rm{L}}}}}{{\bf u}_{\rm{L}}^\dag}} \right) \equiv {{\bf{E'}}_{\rm{L}}}{{\bf{\tilde C}}_{\rm{L}}},\\
\label{eq:E_R_prime_C_R_tilda}
{{\bf{C}}_{\rm{R}}} = {\bf u}_{\rm{R}}^\dag &= \left( { {\bf{Q}}_{{M_{\rm{R}}}}^{ - 1}{{\bf{F}}_{{M_{\rm{R}}}}}} \right)\left( {{\bf{F}}_{{M_{\rm{R}}}}^{-1} {{\bf{Q}}_{{M_{\rm{R}}}}}{{\bf u}_{\rm{R}}^\dag}} \right) \equiv {{\bf{E'}}_{\rm{R}}}{{\bf{\tilde C}}_{\rm{R}}},
\end{align}
\end{subequations}
The transformed input and output matrices $\tilde{\bf B}$ and $\tilde{\bf C}$ can be defined just like Eq.~\eqref{eq:B_matrix} and Eq.~\eqref{eq:C_matrix}.
Instead of computing ${\bf C}{\bf A}^{-1}{\bf B}$, we can compute $\tilde{\bf C} {\bf A}^{-1} \tilde{\bf B}$ with APF using the transformed input/output matrices, after which we undo the transformations using ${\bf{E}}_{\rm{L}}$, ${\bf{E}}_{\rm{R}}$, ${{\bf{E'}}_{\rm{L}}}$, ${{\bf{E'}}_{\rm{R}}}$.
The procedure is illustrated in Fig.~\ref{fig:APF_c}b. 
%The decompression can be carried out efficiently with fast Fourier transforms~\cite{2005_Frigo_FFTW3}.
So far, we have not introduced compression yet; the transformed matrices take up the same sizes as the original matrices, and the transformations can be reversed with no approximation.

While the columns of the original matrix ${\bf B}_{\rm L}$ are spatially extended, those of the transformed matrix ${\bf{\tilde B}}_{\rm{L}}$ can be made spatially localized.
Without the weights ({\it{i.e.}} ${{\bf{Q}}_{{M_{\rm{L}}}}} = {\bf{I}}$), the matrix elements of ${\bf{\tilde B}}_{\rm{L}}$ can be analytically derived from Eq.~\eqref{eq:phi_FD} and Eq.~\eqref{eq:dft} as
\begin{equation}
\label{eq:phi_L_F_L}
{\left( {{{\bf u}_{\rm{L}}}{{\bf F}_{{M_{\rm{L}}}}}} \right)_{ma}} = \sum\nolimits_{b =  - \left( {{M_L} - 1} \right)/2}^{\left( {{M_L} - 1} \right)/2} {{{\left( {{{\bf u}_{\rm{L}}}} \right)}_{mb}}{{\left( {{{\bf F}_{{M_{\rm{L}}}}}} \right)}_{ba}}} =
\begin{cases}
\displaystyle
\frac{{{{\left( { - 1} \right)}^a}}}{{\sqrt {{n_y}{M_{\rm{L}}}} }}\frac{{\sin \left( {{{\pi m{M_{\rm{L}}}}}/{{{n_y}}}} \right)}}{{\sin \left( {\frac{{\pi m}}{{{n_y}}} - \frac{{\pi a}}{{{M_{\rm{L}}}}}} \right)}}, & \displaystyle
\textrm{when } \frac{m}{{{n_y}}} - \frac{a}{{{M_{\rm{L}}}}} \notin \mathbb{Z},\\
\displaystyle
\frac{{{M_{\rm{L}}}}}{{\sqrt {{n_y}{M_{\rm{L}}}} }},\ & \textrm{otherwise},
\end{cases}
\end{equation}
which is the discrete form of the sinc function coming from Fourier transforming a rectangular window ({\it i.e.,} $|k_y| < \omega/c$) in momentum space.
The $a$-th column of ${{\bf{\tilde B}}_{\rm{L}}}$ in Eq.~\eqref{eq:phi_L_F_L} is peaked around a center point $y_{\rm c}^{(a)} \equiv (m_{\rm c}^{(a)}-0.5) \Delta x$ with $m_{\rm c}^{(a)} = {\rm mod}(n_y a / M_{\rm L}, n_y)$, near which its envelope decays as $1/\left| {y - {y_{\rm c}^{(a)}}} \right|$ since $\left|\sin \left( {\frac{{\pi m}}{{{n_y}}} - \frac{{\pi a}}{{{M_{\rm{L}}}}}} \right)\right| \approx \frac{\pi}{n_y}|m-m_{\rm c}^{(a)}|$ in the denominator. Fig.~\ref{fig:phiQL_comp}a-b plot the columns of ${\bf B}_{\rm L}$ and the columns of the transformed matrix ${\bf{\tilde B}}_{\rm{L}}$ with no weight.% Fig.~\ref{fig:phiQL_comp}c shows a column of the transformed matrix ${\bf{\tilde B}}_{\rm{L}}$ without rescaling and decays as $1/\left| {y - {y_{\rm c}^{(a)}}} \right|$.

Given the spatial localization of ${\bf{\tilde B}}_{\rm{L}}$, we can specify a truncation window width $w_{\rm t}$ and set the matrix elements of ${\bf{\tilde B}}_{\rm{L}}$ with $|y-y_{\rm c}^{(a)}| > w_{\rm t}/2$ to zero to make matrix ${\bf{\tilde B}}_{\rm{L}}$ sparse---this is the compression step. Such truncation significantly reduces ${\rm nnz}({\bf{\tilde B}}_{\rm{L}})$, by a factor of $w_{\rm t}/W$. 
There is no need to build the full matrix ${\bf u}_{\rm L}$; we only need to build the elements of ${\bf{\tilde B}}_{\rm{L}}$ within the truncation window $w_{\rm t}$ using Eq.~\eqref{eq:phi_L_F_L}.
There is no need to build the DFT matrix ${{\bf{F}}_{{M_{\rm{L}}}}}$ either, since the transformations can be reversed efficiently with fast Fourier transforms~\cite{2005_Frigo_FFTW3}.

The truncation does introduce small errors since the elements dropped were not exactly zero before.
Fig.~\ref{fig:phiQL_comp}c plots one column of ${\bf{\tilde B}}_{\rm{L}}$ in log scale, which makes the nonzero nature of the $1/\left| {y - {y_{\rm c}^{(a)}}} \right|$ tail more obvious.
To reduce the compression error, we can increase the window size $w_{\rm t}$ and/or make the columns of ${\bf{\tilde B}}_{\rm{L}}$ decay faster.
The relatively slow $1/|y-y_{\rm c}^{(a)}|$ decay in real space comes from the sharp edges of $|k_y| < \omega/c$ in momentum space.
So, we can make ${\bf{\tilde B}}_{\rm{L}}$  more localized using weights that smoothen the sharp edges.
Here, we use the Hann window function~\cite{Smith_1997_book}:
\begin{equation}
\label{eq:cos_weight}
{q^{M}_{a}} = \frac{1}{2}\left[ {1 + \cos \left( {\frac{{2\pi a}}{M}} \right)} \right].
\end{equation}
As Eq.~\eqref{eq:cos_weight} is a superposition of exponential functions, the matrix elements of ${{\bf{\tilde B}}_{\rm{L}}} = {{\bf{u }}_{\rm{L}}}{{\bf{Q}}_{{M_{\rm{L}}}}}{{\bf{F}}_{{M_{\rm{L}}}}}$ can be readily derived as
\begin{equation}
\label{eq:phi_L_Q_L_F_L}
{\left( {{{\bf{u }}_{\rm{L}}}{{\bf Q}_{{M_{\rm{L}}}}}{{\bf F}_{{M_{\rm{L}}}}}} \right)_{ma}} 
= \frac{1}{2}{\left( {{{\bf{u }}_{\rm{L}}}{{{\bf F}}_{{M_{\rm{L}}}}}} \right)_{ma}} + \frac{1}{4}{\left( {{{\bf{u }}_{\rm{L}}}{{{\bf F}}_{{M_{\rm{L}}}}}} \right)_{m,a - 1}} + \frac{1}{4}{\left( {{{\bf{u }}_{\rm{L}}}{{{\bf F}}_{{M_{\rm{L}}}}}} \right)_{m,a + 1}},
\end{equation}
where each term is given by Eq.~\eqref{eq:phi_L_F_L}. Fig.~\ref{fig:phiQL_comp}d-e illustrate this process.
With the Hann window, the columns of ${{\bf{\tilde B}}_{\rm{L}}}$ now decays significantly faster as $1/|y-y_{\rm c}^{(a)}|^3$, shown in Fig.~\ref{fig:phiQL_comp}f. %; a trade-off is that the main peak is now slightly wider.

Additionally, we can also reduce the compression error by padding extra channels.
The $\sin \left( {{{\pi m{M_{\rm{L}}}}}/{{{n_y}}}} \right)$ term in the numerator of Eq.~\eqref{eq:phi_L_F_L} has a width of $(n_y/M_{\rm L})\Delta x$ in $y$.
Therefore, we can pad extra channels to increase $M_{\rm L}$, which reduces the width of the dominant peak, making ${{\bf{\tilde B}}_{\rm{L}}}$ more localized and reducing compression error for the same window size $w_{\rm t}$.
Note that the extra channels we pad do not need to be propagating channels; evanescent ones works equally well since only the transverse profile is involved.
The padded channels can be removed after computing $\tilde{\bf C} {\bf A}^{-1} \tilde{\bf B}$.
In the limit where the padded $M_{\rm L}$ reaches $n_y$, each column of ${{\bf{\tilde B}}_{\rm{L}}}$ will be zero everywhere except at the pixel of $m=m_{\rm c}^{(a)}$, so the compression error completely vanishes even when the truncation window $w_{\rm t}$ is a single-pixel ($\Delta x$) wide; this is equivalent to computing the scattering matrix in an overcomplete spatial basis.

In Sec.~\ref{sec:APF_c_error}, we will give a detailed analysis of the compression error for the mm-wide metasurface examples considered in the main text; by choosing the truncation window size $w_{\rm t}$ and the number of channels to pad, we can make the compression error arbitrarily small.

Lastly, we note that ${{\bf{\tilde B}}_{\rm{L}}}$ in Eq.~\eqref{eq:phi_L_F_L} and Eq.~\eqref{eq:phi_L_Q_L_F_L} is already real-valued, so ${{\bf{\tilde C}}_{\rm{L}}}={{\bf{\tilde B}}_{\rm{L}}}^{\rm T}$ whenever $M_{\rm L} = M_{\rm L}'$.
The channel-index flipping described in Sec.~\ref{sec:symmetrize} no longer necessary for making matrix ${\bf K}$ symmetric.

\section{APF pseudocode \label{sec:pseudocode}}

The pseudocodes of APF and APF-c are shown below, which is also the structure of our implementation made open-source at~\cite{MESTI_GitHub}. 
One can specify an arbitrary system contained in input argument \texttt{syst} (including the permittivity profile, wavelength, discretization grid size, boundary conditions, and PML parameters), arbitrary lists of source profiles given by matrix {\bf B}, and arbitrary output projections given by matrix {\bf C}. 
The algorithm returns the scattering matrix $\bf{S}$.

\smallskip
\smallskip

\begin{algorithm}[H] % Note: the [H] is needed; else it will throw errors because float is not compatible with revtex
\caption{APF}
\begin{algorithmic}
\Require \texttt{syst}, $\bf{B}$, $\bf{C}$, $\bf{D}$ \Comment{\texttt{syst} specifies the system; ${\bf B}$, ${\bf C}$ specify the inputs and outputs; ${\bf{D}} = {\bf{C}}{\bf A}_0^{-1}{\bf{B}} - {\bf S}_0$.}
\Ensure $\bf{S}$
\State $\bf{A} \gets$ \texttt{syst} \Comment{Build the sparse matrix $\bf{A}$ for the Maxwell operator.}
\State ${\bf{K}} = \left[ {{\bf{A}},{\bf{B}};{\bf{C}},{\bf{0}}} \right]$ \Comment{Build the augmented matrix $\bf{K}$.}
\State ${\bf H} = \bf{K}/\bf{A}$ \Comment{Compute the Schur complement $\bf{K}/\bf{A} = -{\bf C}{\bf A}^{-1}{\bf B}$.}
\State $\bf{S} = - {\bf H} - \bf{D}$ \Comment{Scattering matrix $\bf{S} = {\bf C}{\bf A}^{-1}{\bf B} - {\bf D}$.}
\State \textbf{return} $\bf{S}$
\end{algorithmic}
\end{algorithm}
%\quad
\smallskip

\begin{algorithm}[H] % Note: the [H] is needed; else it will throw errors because float is not compatible with revtex
\caption{APF-c}
\begin{algorithmic}
\Require \texttt{syst}, $\bf{B}$, $\bf{C}$, $\bf{D}$ \Comment{\texttt{syst} specifies the system; ${\bf B}$, ${\bf C}$ specify the inputs and outputs; ${\bf{D}} = {\bf{C}}{\bf A}_0^{-1}{\bf{B}} - {\bf S}_0$.}
\Ensure $\bf{S}$
\State $\bf{A} \gets$ \texttt{syst} \Comment{Build the sparse matrix $\bf{A}$ for the Maxwell operator.}
\State ${{\bf{\tilde B}}}, {{\bf{E}}},{{\bf{\tilde C}}}, {{\bf{E'}}}  \gets \bf{B}, \bf{C}$ \Comment{Compress the matrices  ${{\bf{B}}}$ and ${{\bf{C}}}$.}
\State $\bf{\tilde K} = [\bf{A}, {{\bf{\tilde B}}}; {{\bf{\tilde C}}},{\bf{0}}]$ \Comment{Build the augmented matrix $\bf{\tilde K}$.}
\State ${\bf \tilde{H}} = {\bf \tilde{K}}/\bf{A}$ \Comment{Compute the Schur complement ${\bf \tilde{K}}/\bf{A} = -{\bf \tilde{C}}{\bf A}^{-1}{\bf \tilde{B}}$.}
\State $\bf{S} = - {\bf E}'{\bf \tilde{H}}{\bf E} - \bf{D}$ 
\Comment{Decompress with the matrices ${{\bf{E}}}$ and ${{\bf{E'}}}$.}
\State \textbf{return} ${\bf S}$
\end{algorithmic}
\end{algorithm}

\section{System-size scaling for disordered media simulations\label{sec:scaling_fixed_aspect_ratio}}

\begin{figure*}[t]
\includegraphics[width=0.9\textwidth]{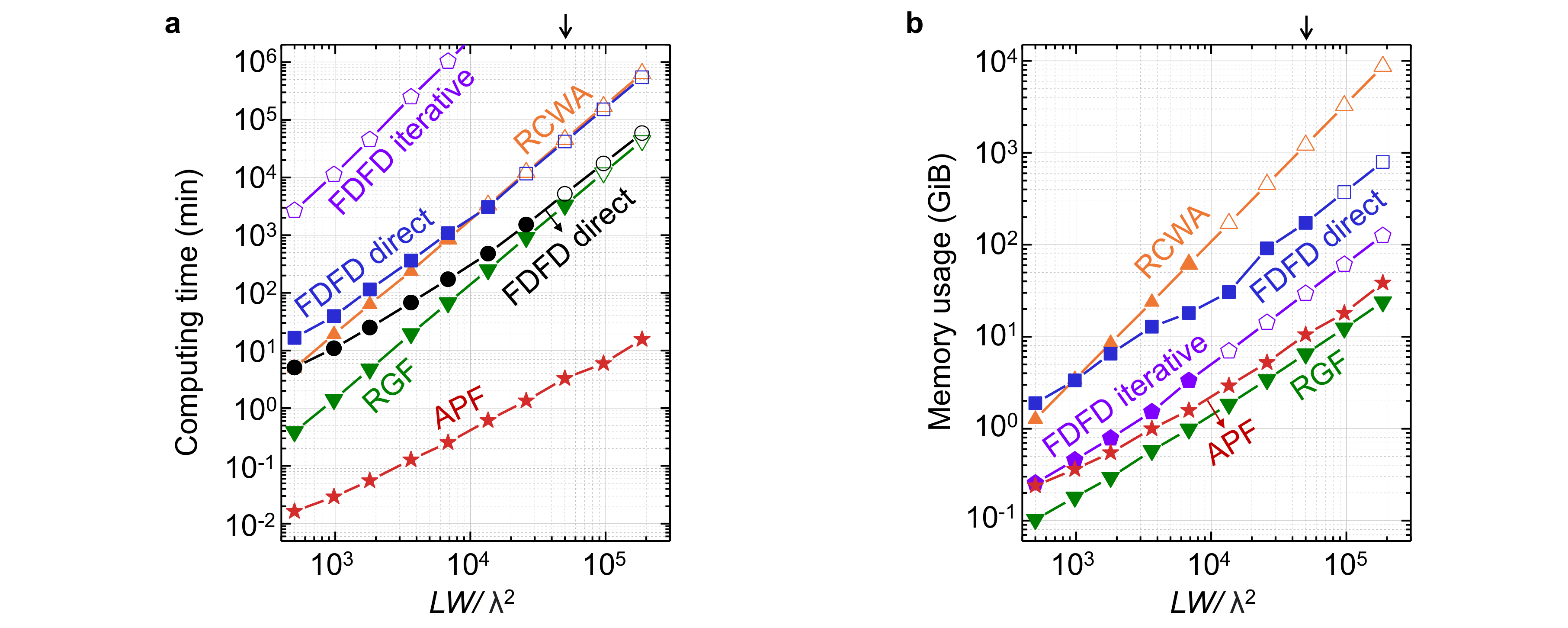}
\caption{\label{fig:scaling_time_mem} {\bf Scaling of computing time and memory usage using different methods}. Computing time ({\bf{a}}) and memory usage ({\bf{b}}) versus the system size $LW/\lambda^2$.
The two ``FDFD direct'' curves correspond to an unmodified version of MaxwellFDFD (blue curve) and one modified to have the LU factors reused for different inputs (black curve).
The arrows on top indicate the system considered in Fig.~2 of the main text.}
\end{figure*}

Figure~\ref{fig:scaling_time_mem} shows how the computing time and memory usage scale with the system size using different methods, for the disordered system in Fig.~2 of the main text with the aspect ratio $W/L = 5$ fixed while the overall system size ($W$ and $L$) varies.
Some methods require more computing resources than we have access to, so their data points (open symbols) are extrapolated from smaller number of input angles $M$ and/or smaller systems.
We note that among all of these methods, APF exhibits the best scaling both in terms of computing time and in terms of memory usage, and is many orders of magnitude faster for large systems.

\begin{figure*}[b]
\includegraphics[width=0.85\textwidth]{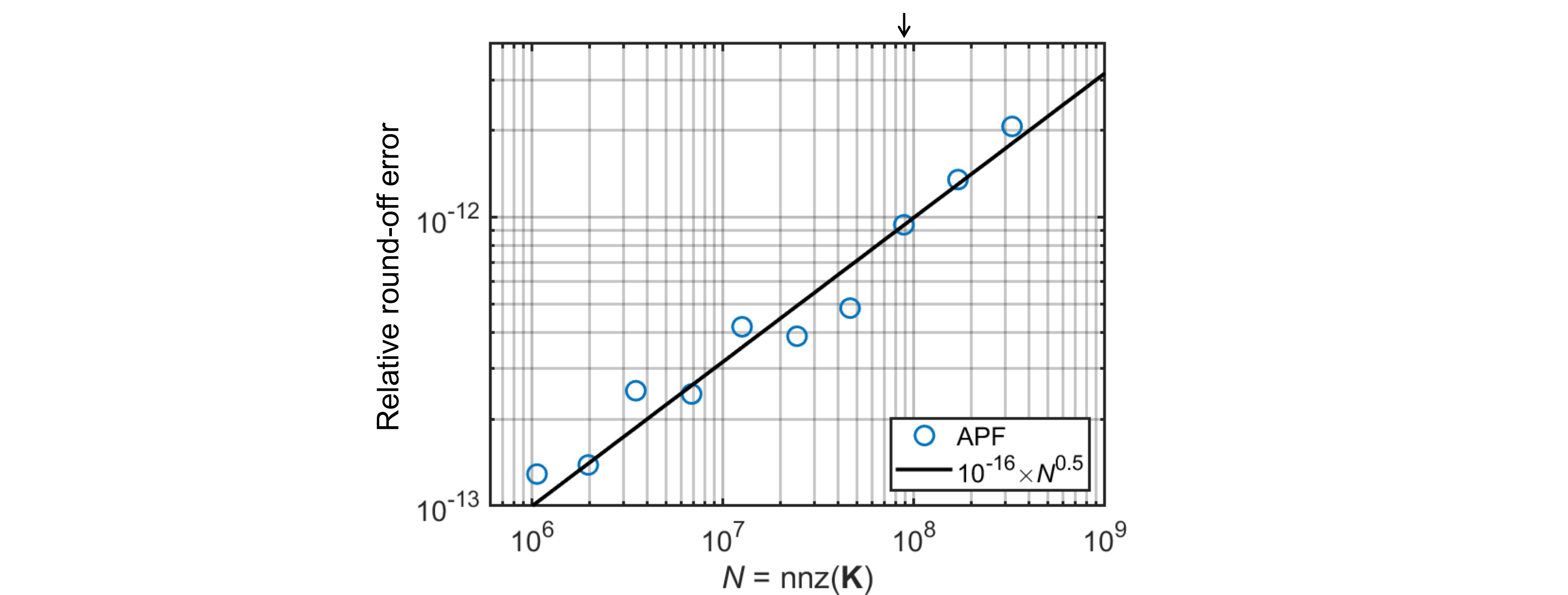}
\caption{\label{fig:error_APF} {\bf Round-off error of APF}. Circles show the relative difference between scattering matrices computed using APF and those computed using direct method with iterative refinement that iterates until machine precision is reached. Black line is a fitting curve. The black arrow on top indicates the system considered in Fig.~2 of the main text.}
\end{figure*}

\section{Round-off error of APF}

While APF is in theory exact aside from discretization error (and compression error if APF-c is used), numerical round off is also present, and such round-off errors often grow with the system size.
Therefore, it is necessary to check if the round-off error of APF is within the acceptable range.
To characterize the round-off error, we compare results from APF to those obtained from a standard direct solver but with additional iterative refinement~\cite{1989_Arioli_SIAM} steps that iterate until the entire solution (with all input channels) reaches machine-precision accuracy.
Double precision is used through out.

From that, we evaluate the relative $\ell^2$-norm error, ${\left\| S_{\rm{APF}} - S_0 \right\|_2}/{\left\| S_0 \right\|_2}$, for the systems in Sec.~\ref{sec:scaling_fixed_aspect_ratio}, where $S_0$ is the scattering matrix (reshaped into a vector) computed with iterative refinement, and $S_{\rm{APF}}$ is that from APF. The difference between the two is the round-off error of APF. The relative round-off error with respect to $N$ = nnz({\bf{K}}) scales as $\mathcal{O}(N^{0.5})$ (Fig.~\ref{fig:error_APF}); it is only $10^{-12}$ for the system considered in Fig.~2 of the main text where $N \approx 10^{8}$, and is estimated to be only $10^{-10}$ even for an extremely large system with a trillion matrix elements. We therefore conclude that the round-off error of APF is negligible even for the largest system one would possibly simulate. 

\section{Metalens design and simulation \label{sec:design_meta}} 

\begin{figure*}[t]
\includegraphics[width=1.0\textwidth]{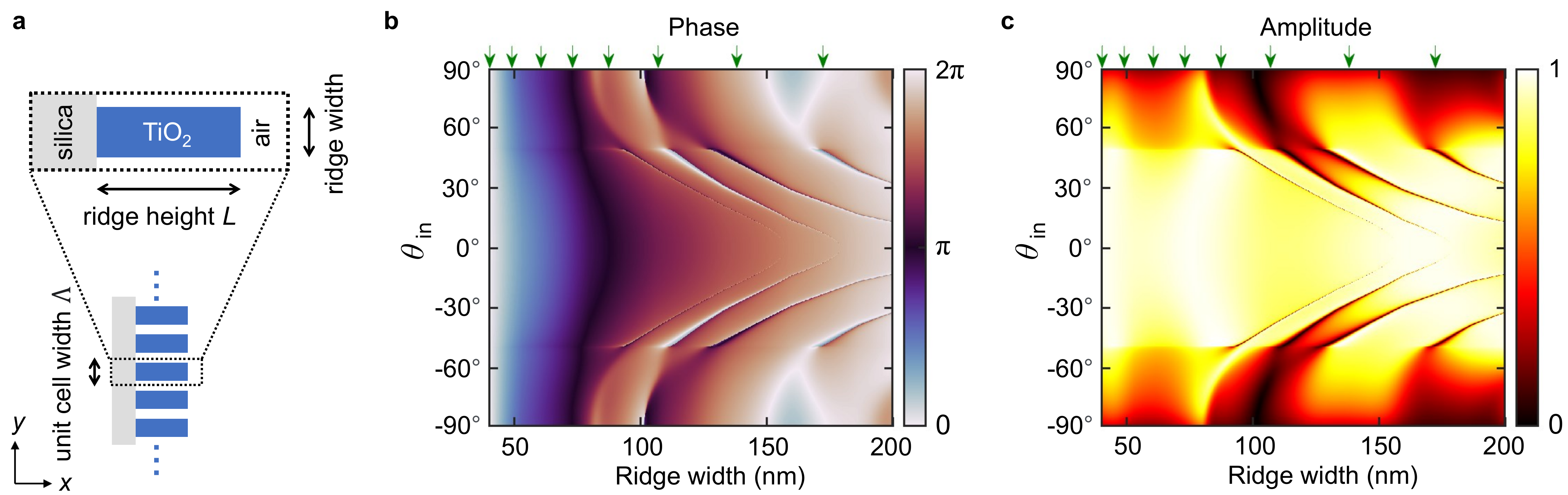}
\caption{\label{fig:metalens_design} {\bf Unit cells of the metasurface}. {\bf{a}}, Schematic structure of a periodic array of unit cells considered here. One unit cell is simulated with Bloch periodic boundary condition in $y$. {\bf{b-c}}, Phase ({\bf{b}}) and amplitude ({\bf{c}}) maps of the zeroth-order transmission coefficient for different ridge widths and incident angles. The green arrows on top indicate the ridge widths used for the design.}
\end{figure*}

\begin{table}[b]
\begin{tabularx}{0.8\textwidth}{ c *{8}{Y} }
%{|c||c|c|c|c|c|c|c|c|}
%\hline
\toprule
\,\,\,Relative phase shift\,\,\, & 0  & $\pi$/4  & $\pi$/2  & 3$\pi$/4  & $\pi$  & 5$\pi$/4   & 3$\pi$/2 & $7\pi/4$   \\
%\hline
\midrule
Ridge width (nm)    & 40.0 & 49.1 & 60.7 & 73.1 & 87.4 & 107.1 & 138.2 & 172.3 \\ 
%\hline
\bottomrule
\end{tabularx}
\caption{\label{tab:metalens_design} {\bf Ridge widths used for the metalens design}. Eight ridge widths and the corresponding relative transmission phase shifts at normal incidence.
}
\end{table}

We start by describing how the hyperbolic and quadratic metalenses in Figs.~4--5 of the main text are designed. The metalens operates at wavelength $\lambda = 532$ nm and consists of 4,178 unit cells.
Each unit cell ({\it i.e}, a meta-atom) has a titanium dioxide (TiO$_2$) ridge (refractive index $n = 2.43$) sitting on a silica substrate ($n = 1.46$) in air ($n=1$), as shown in Fig.~\ref{fig:metalens_design}a. The ridge height in $x$ is fixed at $L=$ 600 nm, and the width $\Lambda$ of an unit cell is fixed at 239.4 nm. We consider ridge widths between 45 nm and 200 nm.
We use a grid size of $\Delta x = \lambda/40$ for the finite-difference discretization, which ensures that the transmission phase shifts are accurate to within 0.1 radian (see Sec.~\ref{sec:discretization_error}).
We then map out the phase and amplitude of the zeroth-order ({\it i.e.}, $a=b=0$) transmission coefficient of the unit cell with Bloch periodic boundary condition in $y$, for different ridge widths and different incident angles, as shown in Fig.~\ref{fig:metalens_design}b--c. 
From these results, we pick eight ridge widths as indicated by the green arrows in Fig.~\ref{fig:metalens_design}b--c and summarized in Table~\ref{tab:metalens_design}, which provide eight equally-spaced transmission phase shifts covering 0 to 2$\pi$ at normal incidence.

For a hyperbolic metalens, the transmission phase shift at normal incidence should be space-dependent with a hyperbolic profile~\cite{aieta2012aberration}
\begin{equation}
\label{eq:phi_hyperbolic}
\Phi_{\rm{hyperbolic}} \left( y \right) =  - \frac{{2\pi }}{\lambda }\sqrt {{f^2} + {y^2}},
\end{equation}
where $f$ is the focal length.
The coordinate $y$ here is zero at the center of the metalens. A hyperbolic metalens can achieve diffraction-limited focusing at normal incidence but comes with off-axis aberrations at oblique incidence~\cite{aieta2013aberrations}.
For a quadratic metalens, the transmission phase shift should follow a quadratic profile~\cite{Pu2017_OE}
\begin{equation}
\Phi_{\rm{quadratic}} \left( y \right) =  - \frac{{2\pi }}{\lambda }\frac{{{y^2}}}{{2f}}.
\end{equation}
We construct the metalenses from the eight unit cells in Table~\ref{tab:metalens_design}, with the ridge width of each unit cell chosen based on the desired normal-incidence transmission phase shift at the center of that unit cell.
The metalenses consists of 4,178 unit cells, with an overall width of $W=1000.2$ {\textmu}m. For the hyperbolic metalens, we use a focal length of $f= 300$ {\textmu}m (corresponding to numerical aperture NA = 0.86). For quadratic metalenses, there is a maximal effective numerical aperture of   ${\rm{N}}{{\rm{A}}_{e{\rm{ff}}}} = {n_{\rm{t}}}/\sqrt 2$ where $n_{\rm{t}}$ is the refractive index of the medium on the transmitted side~\cite{Lassalle2021_ACSP}, so we use a larger focal length of $f= 500$ {\textmu}m (corresponding to NA = 0.71).

Such design, although standard, is quite simplistic as it only considers normal incidence and assumes that the unit-cell simulations (which are carried out for fully periodic systems) are sufficient for capturing the response of the aperiodic metalens. The purpose of this design here is not to realize a high-performance metalens, but simply to provide a test system to benchmark the APF method.

After building $\varepsilon_{\rm r}(x,y)$ of the mm-wide metalens, we carry out full-wave simulations with APF-c to compute its transmission matrix.
The simulation domain is schematically illustrated in Fig.~4a of the main text, with PML on all four sides to describe a fully open system; also see the Methods section of the main text.
The pre-compression inputs are line sources in the silica substrate immediately behind the TiO$_2$ ridges, which generate incident plane waves truncated with a rectangular window over the width $W$ of the metalens; this models the effect of having an aperture that blocks incident light beyond width $W_{\rm in} = W$.
We use the set of propagating input channels for a periodic boundary in $y$ with width $W_{\rm in}$, and restrict to the $2W_{\rm in}/\lambda =$ 3,761 incident angles with $|\theta_{\rm in}^{\rm substrate}| \le {\rm asin}(1/n_{\rm substrate}) = 43^\circ$ (namely, 
$|\theta_{\rm in}| \le 90^\circ$ in air);
% = |\sin^{-1}(n_{\rm substrate}\sin\theta_{\rm in}^{\rm substrate})|
as described in Sec.~\ref{sec:chan_homo_sp}, this is the minimal number of channels we need to specify wavefronts incident from air prior to entering the substrate. % at the Nyquist--Shannon rate.
%As the thickness $L=0.6$ $\mu$m is small, there is little transmitted light beyond the width of the metalens.
For the output projections, we take the total field $E_z(x=L,y)$ immediately after the TiO$_2$ ridges across a width $W_{\rm out} = W + 40\lambda$ in $y$ that is sufficiently large to contain all of the transmitted light, 
and project it onto the $2W_{\rm out}/\lambda =$ 3,841 propagating plane waves in air with $|\theta_{\rm out}| \le 90^\circ$;
this is the minimal number of output projections we need to specify the propagating components of the transmitted light.
The input and output matrices are compressed for the computations, following the steps in Sec.~\ref{sec:APF_c}.

\section{System-size scaling for metasurface simulations with RCWA and RGF}

\begin{figure*}[t]
\includegraphics[width=0.85\textwidth]{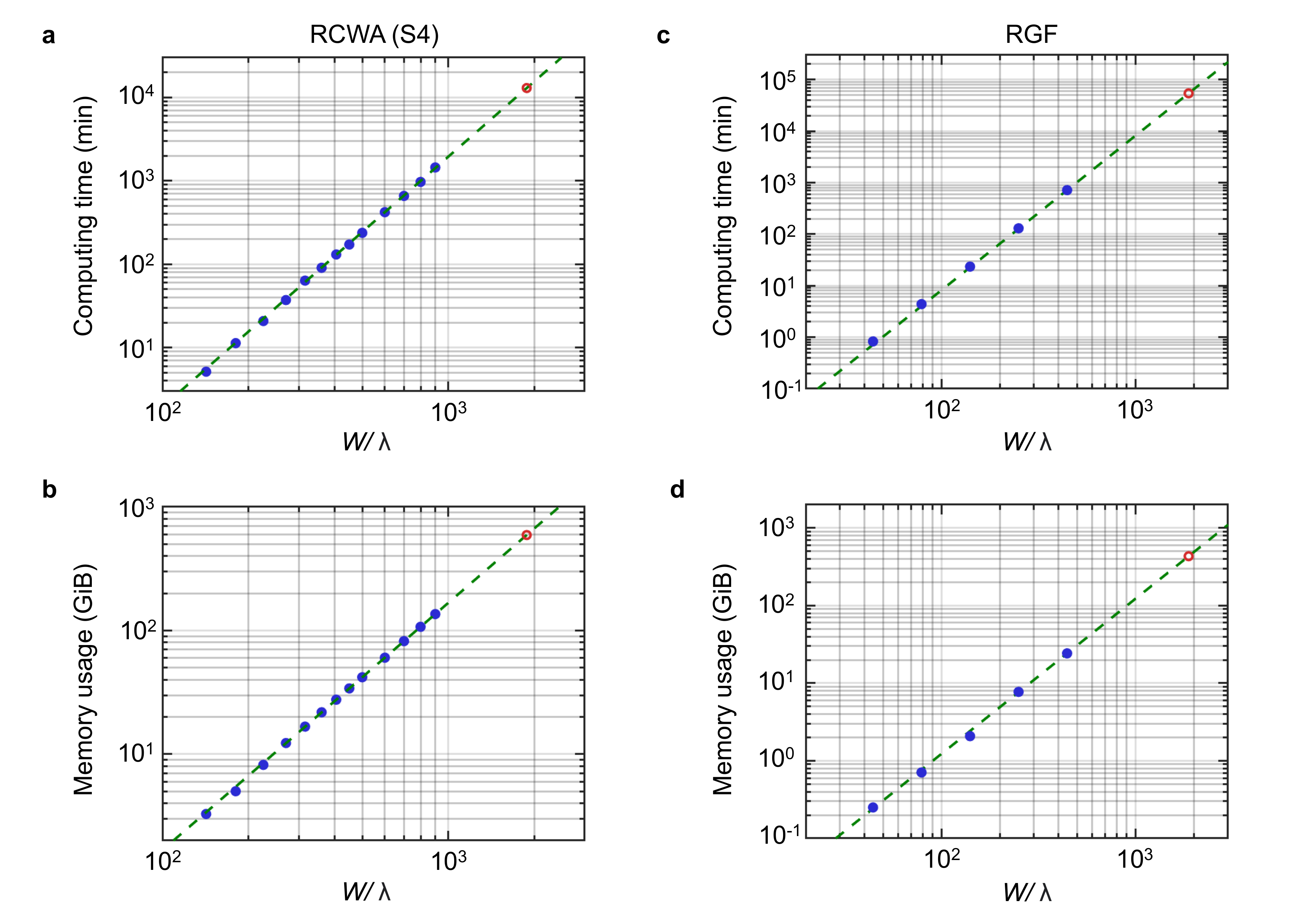}
\caption{\label{fig:extra_RGF_RCWA} {\bf System-size scaling for metasurface simulations with RCWA and RGF}. {\bf{a-b}}, Computing time and memory usage of RCWA to simulate metasurfaces with varying widths $W$. {\bf{c-d}}, Corresponding results with RGF. Blue circles are from simulations, green dashed lines are $\mathcal{O}(W^3)$ and $\mathcal{O}(W^2)$ fitting curves, and red circles are the points we extrapolate to to estimate the computing time and memory usage for simulating a mm-wide metasurface, as shown in Fig.~4 of the main text.}
\end{figure*}

As RCWA and RGF work with dense matrices, they do not scale well with the system size, and simulating the mm-wide metasurface requires more computing resources than we have access to.
Therefore, we extrapolate from smaller systems.
We fix the thickness of the metasurface, and run RCWA and RGF simulations with increasing metasurface width $W$;
the computing time and memory usage are shown as blue circles in Fig.~\ref{fig:extra_RGF_RCWA}.
For both methods, the computing time scales as $\mathcal{O}(W^3)$, which is the time scaling to invert and to multiply size-$\mathcal{O}(W)$ dense matrices; we fit the data with $\mathcal{O}(W^3)$ curves (green dashed lines) and extrapolate to a mm-wide metasurface (red circle).
Similarly, the memory usages scale as $\mathcal{O}(W^2)$, which we use for extrapolation.

As described in the Methods section of the main text, the RCWA and RGF simulations here adopt a periodic boundary condition in $y$ with width $W$ (not $W_{\rm out}$), without free space and without PML. So, the numbers here are a lower bound of the would-be numbers when an open boundary in $y$ is implemented.

\section{Angular spectrum propagation \label{sec:ASP}} 

We use angular spectrum propagation (ASP) to obtain field profile in the free space after the metalens.
We express ${E_z}\left( {x,y} \right)$ in terms of its Fourier components ${\tilde E_z}\left( {x,{k_y}} \right)$ as
%Consider the total field ${E_z}\left( {x = L,y} \right)$ in air, in the plane $x = L$ immediately after the metasurface. Let the function ${\tilde E_z}\left( {x = L ;{k_y}} \right)$ represent the angular spectrum of ${E_z}\left( {x = L,y} \right)$,
\begin{equation}
\label{eq:ASP_IFT}
%{\tilde E_z}\left( {x;{k_y}} \right) = \frac{1}{{\sqrt {2\pi } }} \int_{ - \infty}^{\infty} {{E_z}\left( {x,y} \right){e^{ - i{k_y}y}}dy},
{E_z}\left( {x,y} \right)
= \frac{1}{{\sqrt {2\pi } }} \int_{ - \infty}^{\infty} d k_y \, {\tilde E_z}\left( {x,{k_y}} \right) {e^{ i{k_y}y}}.
\end{equation}
Plugging Eq.~\eqref{eq:ASP_IFT} into the source-free Eq.~\eqref{eq:A_psi} with $x \ge L$ [where $\varepsilon_{\rm{r}}(x,y)=1$]
gives $\frac{\partial^2}{\partial x^2} {\tilde E_z}\left( {x,{k_y}} \right) = -k_x^2$ with $k_x = \sqrt{(\omega/c)^2-k_y^2}$.
As there is no light incident from the right, light must propagate or decay to the right, so
\begin{equation}
\label{eq:ASP_propagate}
{\tilde E_z}\left( {x\ge L,{k_y}} \right) = 
{\tilde E_z}\left( {x=L,{k_y}} \right)e^{i k_x (x-L)}.
\end{equation}
Therefore, given the total field ${E_z}\left( {x = L,y} \right)$ immediately after the metasurface, we can take its Fourier transform to obtain ${\tilde E_z}\left( {x=L,{k_y}} \right)$, propagate it forward with Eq.~\eqref{eq:ASP_propagate}, and obtain the field profile anywhere in the free space after the metalens with Eq.~\eqref{eq:ASP_IFT}.
This method is called ``angular spectrum propagation'' (ASP)~\cite{2017_Goodman_book}.

The evanescent components (for which $|k_y| > \omega/c$) decay exponentially.
Since we are not interested in the near-field close to the metasurface, we can ignore the evanescent components, and replace the $\int_{ - \infty}^{\infty} d k_y$ integration range with $\int_{ -\omega/c}^{\omega/c} d k_y$.
Therefore, to perform ASP, we only need the propagating Fourier components of ${E_z}\left( {x = L,y} \right)$, which are precisely what's contained in the transmission matrix we computed with APF-c.

There is one subtlety.
In practice, the continuous integration over $k_y$ must be approximated with a discrete summation.
The transmission matrix elements have the transverse wave number $k_y$ of the transmitted light evenly spaced by $2\pi/W_{\rm out}$ as described below Eq.~\eqref{eq:phi_a}.
If we perform ASP and the approximated integration directly with such $2\pi/W_{\rm out}$ momentum spacing, it will introduce an artificial periodic boundary with periodicity $W_{\rm out}$; light that reaches the boundary will wrap around and reenter from the other side instead of leaving the domain of interest.
Therefore, we need a transverse momentum spacing finer than $2\pi/W_{\rm out}$.
To achieve so, we first evaluate ${E_z}\left( {x = L,y} \right)$ using Eq.~\eqref{eq:psi_tot_outside_continuous}, restricting the summation over $b$ to the propagating components contained in the computed transmission matrix.
Then, we evaluate its Fourier components
\begin{equation}
\label{eq:ASP_FT}
{\tilde E_z}\left( {x=L,{k_y}} \right) = \frac{1}{{\sqrt {2\pi } }} \int_{ - \infty}^{\infty} dy\, {{E_z}\left( {x=L,y} \right){e^{ - i{k_y}y}}}
\end{equation}
on a finer grid of transverse momentum $k_y$ spaced by $2\pi/W_{\rm ASP}$.
Here, we use $W_{\rm ASP} \approx 2 W$, which is sufficient for eliminating the periodic wrapping artifacts within the domain of interest ($x \lesssim f$, $|y|<W/2$).

The integration range $\int_{ - \infty}^{\infty} dy$ in Eq.~\eqref{eq:ASP_FT} can be truncated to the width $W_{\rm out}$ where we perform the output projections, since the field beyond such window is negligibly small.
A high resolution of $\Delta x = \lambda/40$ is not necessary for this spatial integration since the refractive index is lower ($n=1$ in air) and since only the propagating components (which vary slowly in $y$) are of interest, so we use a coarser resolution of $\Delta x' = \lambda/8$ for the integration in Eq.~\eqref{eq:ASP_FT}.
The evaluations of Eq.~\eqref{eq:psi_tot_outside_continuous}, Eq.~\eqref{eq:ASP_FT}, and then Eq.~\eqref{eq:ASP_IFT} can all be done efficiently with fast Fourier transforms~\cite{2005_Frigo_FFTW3}.

ASP is an exact method;
when the continuous integrals are not replaced by discrete summations, ASP is mathematically equivalent to the Rayleigh--Sommerfeld diffraction integral~\cite{2022_Cubillos_ANM} used in Ref.~\cite{Pestourie2018_OE} where it was referred to as the equivalent-current formulation.
%But numerically, ASP is more efficient thanks to fast Fourier transforms.
%ASP is efficient and simple to implement
%, but when $x \gg W$, one may consider using more sophisticated methods~\cite{2022_Cubillos_ANM} to avoid the large $W_{\rm ASP}$.

\section{Metalens transmission efficiency and Strehl ratio} 

The transmission efficiency and the Strehl ratio are important metrics for assessing the performance of a metalens.
Here we evaluate the incident-angle dependence of these quantities using the transmission matrix computed with APF-c.
We define the transmission efficiency as the total transmitted flux in $x$ direction divided by the total incident flux in $x$ direction.
Since our definition of the transmission matrix [as in Eq.~\eqref{eq:psi_tot_outside_continuous}] is already flux-normalized, the transmission efficiency of a truncated incident plane wave with angle $\theta_{\rm in}^{(a)}$ is simply
\begin{equation}
\label{eq:trans_eff}
T_a = \sum\limits_{b} {{{\left| {{t_{ba}}} \right|}^2}}.
\end{equation}

The Strehl ratio is defined as the actual intensity at the focal spot $|E_z(x=L+f, y=f_{\rm f}|^2$ divided by the would-be intensity $|E_z^{\rm (ideal)}({\bf r}_{\rm f})|^2$ at the focus if perfect diffraction-limited focusing were achieved with the given transmission efficiency.
Since the focus location $y_f$ depends on the incident angle and is not clearly defined at large angles where aberrations are significant, we evaluate $|E_z(x=L+f, y=f_{\rm f}|^2$ as $\max_{y} |E_z(x=L+f, y|^2$.
With the transmission efficiency adjusted, the Strehl ratio is therefore
\begin{equation}
{\rm{SR}}(\theta_{\rm in}^{(a)}) = \frac{{\mathop {\max }\limits_y  {{{\left| {E_z^{(a)}( {x = L + f,y})} \right|}^2}} }\big/T_a}{|E_z^{\rm (ideal)}({\bf r}_{\rm f})|^2/T_{\rm ideal}},
\end{equation}
where the field profile $E_z^{(a)}( {x = L + f,y})$ with incident angle $\theta_{\rm in}^{(a)}$ is computed with angular spectrum propagation as described in Sec.~\ref{sec:ASP}.
To compute $|E_z^{\rm (ideal)}({\bf r}_{\rm f})|^2/T_{\rm ideal}$, we let the ideal field profile immediately after the metalens be $E_z^{\rm (ideal)}(x=L,y) = e^{i\Phi_{\rm{hyperbolic}}(y)}$ as in Eq.~\ref{eq:phi_hyperbolic};
%, project it onto the propagating waves as in Eq.~\eqref{eq:psi_tot_outside_continuous} to get the transmission coefficients, and compute its transmission efficiency $T_{\rm ideal}$ with Eq.~\eqref{eq:trans_eff}
its associated transmission efficiency $T_{\rm ideal}$ and field at the focus $E_z^{\rm (ideal)}({\bf r}_{\rm f}) = E_z^{\rm (ideal)}(x=L+f, y=0)$ are then evaluated with the same procedure as above. %; its relative amplitude is not important due to the normalization.

\section{Locally periodic approximations}

Here we consider the locally periodic approximation (LPA), which is commonly used when full-wave simulations of the entire metasurface take too much computing resources.
As the field $E_z^{(a)}(x = L,y)$ immediately after the metasurface is the only input for the transmission efficiency and angular spectrum propagation, here we use LPA to approximate $E_z^{(a)}(x = L,y)$.
We consider two LPA formalisms, referred to as LPA I and LPA II here.
In both formalisms, the approach is to divide the metasurface into individual unit cells and assume that the response of each unit cell can be described by the unit-cell simulations in Sec.~\ref{sec:design_meta} and Fig.~\ref{fig:metalens_design} performed for an individual unit cell under Bloch periodic boundary condition.
The unit-cell simulations have incident fields
$E_z^{{\rm in},p}(x,y) = \frac{E_0}{\sqrt{\Lambda k_x^{(a, {\rm L})}}}
\exp\left[i k_x^{(a, {\rm L})} x + i k_y^{(a, {\rm L})} (y-y_p) \right]$
where $y_p = (p-1)\Lambda$ is the starting position of the $p$-th unit cell,
while for the full-metasurface simulation we want the incident field to be
$E_z^{{\rm in}}(x,y) = \frac{E_0}{\sqrt{W k_x^{(a, {\rm L})}}}
\exp\left[i k_x^{(a, {\rm L})} x + i k_y^{(a, {\rm L})} y \right]$,
so a prefactor of $\sqrt{\Lambda/W} \exp\left[i k_y^{(a, {\rm L})} y_p \right]$
needs to be added.

Let $E_z^{(a),p}(x,y)$ with $0\le y \le \Lambda$ be the total field from simulation of the $p$-th unit cell with incident angle $\theta_a$.
Then, LPA II simply stitches together such unit-cell field profiles to approximate $E_z^{(a)}(x = L,y)$, as
\begin{equation}
E_z^{\left( a \right){\textrm{LPA-II}}}\left( {x = L,y} \right) = \sqrt{\frac{{ \Lambda  }}{{ W }}} \sum\limits_{p = 1}^{{N_p}} { {e^{i k_y^{(a, {\rm L})} y_p }} E_z^{(a),p}(x=L,y-y_p)} \Pi_p(y),
\end{equation}
where $N_p =$ 4,178 is the unit of unit cells, and $\Pi_p(y)$ is a rectangular function that equals 1 when $y_p \le y \le y_p+\Lambda$, 0 elsewhere.
LPA II includes all propagating and evanescent components of the unit-cell simulations, all of which are contained in $E_z^{(a),p}(x,y)$.

Oftentimes, only the zeroth-order ({\it i.e.}, $a=b=0$) transmission coefficient of the unit cell is considered.
Therefore, LPA I uses
\begin{equation}
E_z^{\left( a \right){\textrm{LPA-I}}}\left( {x = L,y} \right) = \sqrt{\frac{{ \Lambda  }}{{ W }}} \sum\limits_{p = 1}^{{N_p}} { {e^{i k_y^{(a, {\rm L})} y_p }} E_z^{(a),p; {\rm prop}}(x=L,y-y_p)} \Pi_p(y),
\end{equation}
where 
\begin{equation}
E_z^{(a),p; {\rm prop}}(x=L,y) = t_p \frac{e^{i k_y^{(a, {\rm R})} y }}{\sqrt{\Lambda k_x^{(a, {\rm R})}}}, \quad 0\le y \le \Lambda
\end{equation}
is the zeroth-order propagating component of the $p$-th unit cell as in Eq.~\eqref{eq:psi_tot_outside_continuous},
with $t_p$ being the corresponding transmission coefficient.
Since $\Lambda < \lambda/2$ here, this zeroth-order component is the only propagating component on the transmitted side.

Given the approximate $E_z^{(a)}(x = L,y)$, we then use the same angular spectrum propagation procedure to obtain the approximate $E_z^{(a)}(x = L+f,y)$.

\begin{figure*}[t]
\includegraphics[width=0.85\textwidth]{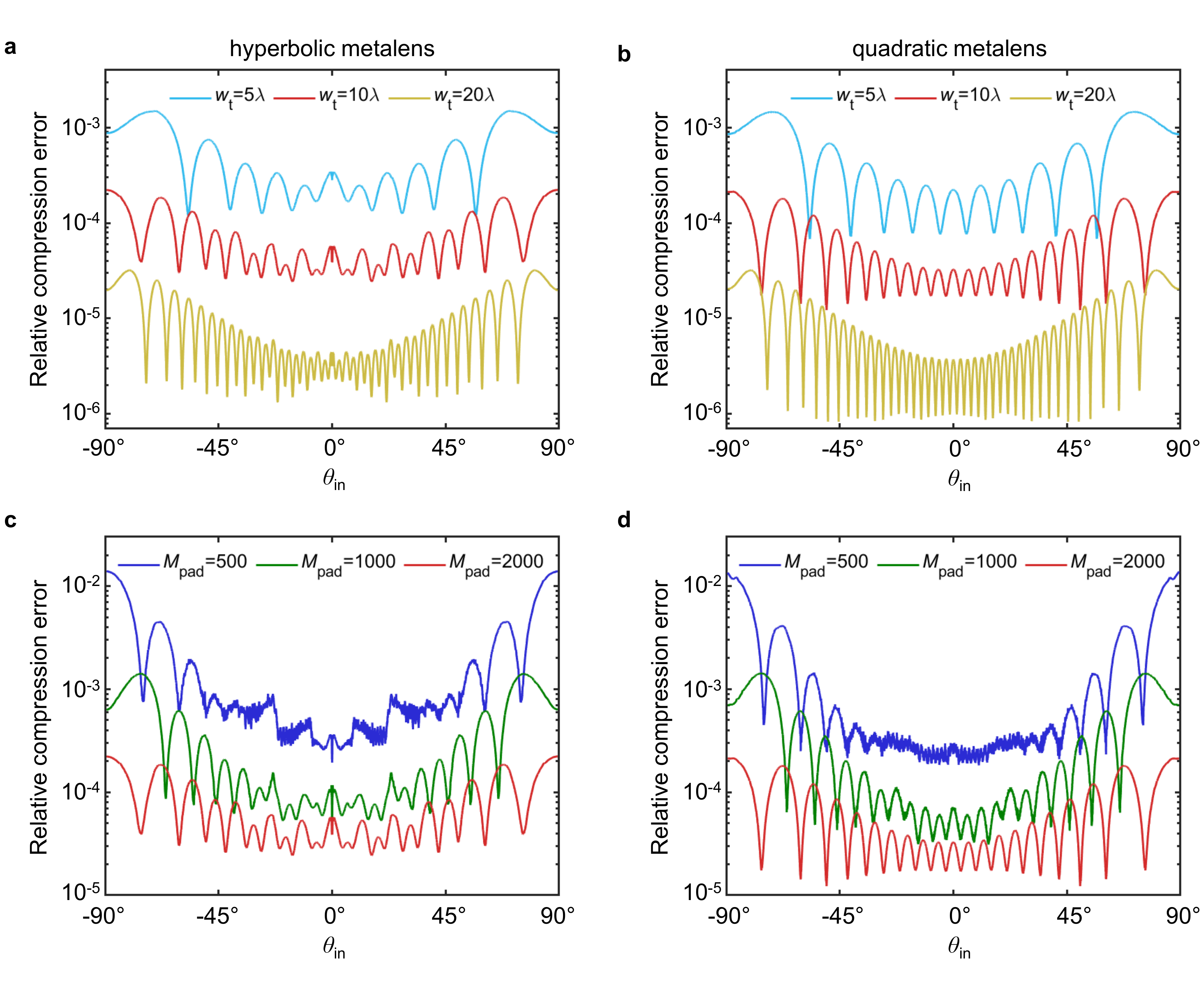}
\caption{\label{fig:APF_c_char} {\bf Compression error of APF-c}. 
{\bf{a--b}}, Angle dependence of the relative compression error for different truncation window widths $w_{\rm{t}}$ for the mm-wide hyperbolic metalens ({\bf a}) and quadratic metalens ({\bf b}), with the number of padded channels fixed at $M_{\rm{pad}} =$ 2,000. 
{\bf{c--d}}, Compression error for different choices of $M_{\rm{pad}}$ with the truncation window width fixed at $w_{\rm{t}} = 10 \lambda$.
The red curves, with $w_{\rm{t}} = 10 \lambda$ and $M_{\rm{pad}} =$ 2,000, correspond to the choice used for Figs.~4--5 in the main text.
}
\end{figure*}

\section{APF-c compression error \label{sec:APF_c_error}} 

As described in Sec.~\ref{sec:APF_c}, the APF-c compression error can be reduced to arbitrarily small by increasing the truncation window $w_{\rm{t}}$ and/or by padding $M_{\rm{pad}}$ additional channels.
Here, we characterize the APF-c compression error of the metalens systems. As described in the main text, we compute the relative $\ell^2$-norm error ${\left\| I - I_0 \right\|_2}/{\left\| I_0 \right\|_2}$, with $I_0$ being a vector containing the intensity $|E_z^{(a)}(x = L+f,y)|^2$ at the focal plane within $|y|<W/2$ calculated from APF without compression, and $I$ from APF-c. Fig.~\ref{fig:APF_c_char} plots the error as a function of the incident angle for varying $w_{\rm{t}}$ and $M_{\rm{pad}}$.
The Hann window is used, and the same $w_{\rm{t}}$ and $M_{\rm{pad}}$ are used on both the left (incident) and the right (transmitted) sides.

\begin{figure*}[t]
\includegraphics[width=0.88\textwidth]{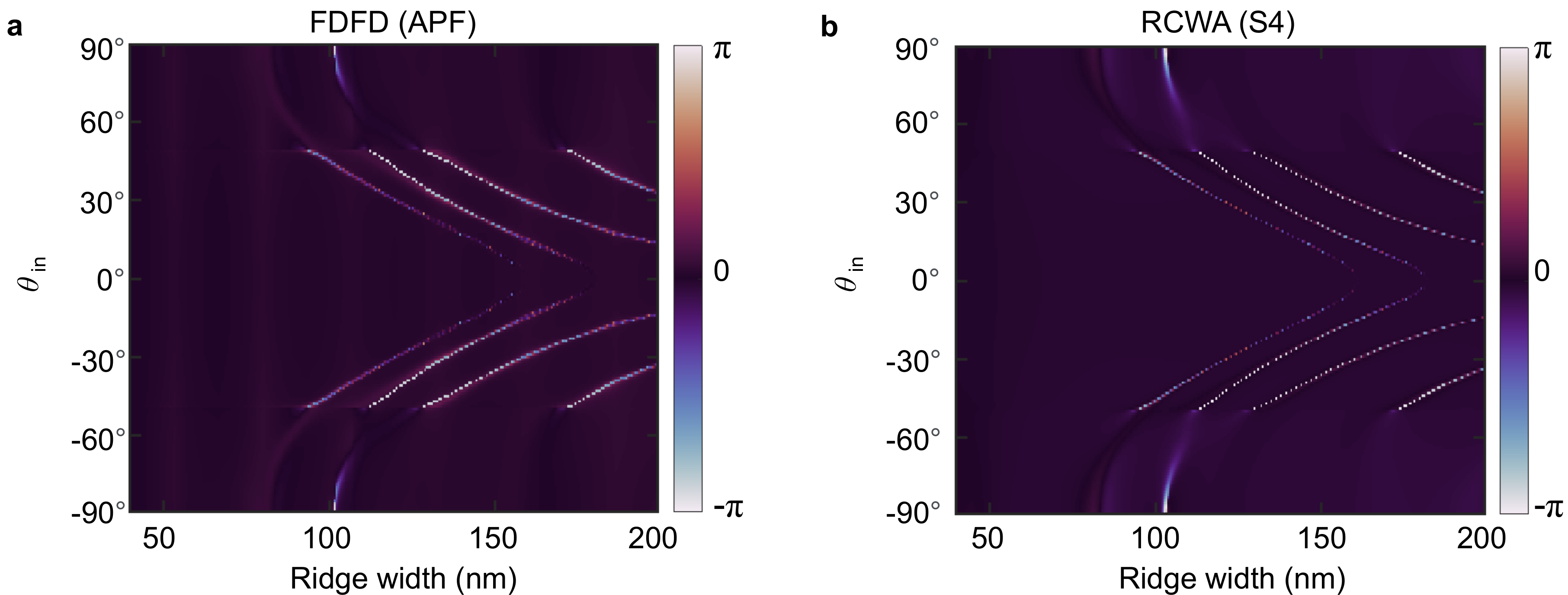}
\caption{\label{fig:dis_error} {\bf Discretization error}. Discretization error of ({\bf{a}}) FDFD, characterized by $\Phi_{\rm{FDFD}}^{\Delta x = \lambda/40}  - {\Phi_{\rm{FDFD}}^{\Delta x = \lambda/240}}$, and ({\bf{b}}) RCWA, characterized by $\Phi_{\rm{RCWA}}^{{\rm numG}=5}  - {\Phi_{\rm{RCWA}}^{{\rm numG}=99}}$,
for the meta-atom zeroth-order transmission phase shifts.
}
\end{figure*}

\section{Discretization error\label{sec:discretization_error}} 
The grid size $\Delta x$ in the finite-difference discretization and the number of Fourier components in RCWA are chosen to ensure sufficient accuracy, which we describe here.

For metalenses, the most important property is the transmission phase shift. Figure~\ref{fig:dis_error}a shows the discretization error of the zeroth-order transmission phase shift, $\Phi_{\rm{FDFD}}^{\Delta x = \lambda/40}  - {\Phi_{\rm{FDFD}}^{\Delta x = \lambda/240}}$, of the unit cells described in Sec.~\ref{sec:design_meta}.
We see that the discretization error is negligible away from the resonances, and the error at the resonances arises because the angle at which the resonance exists is highly sensitive on the structure.
Here, the wrapped  $|\Phi_{\rm{FDFD}}^{\Delta x = \lambda/40}  - {\Phi_{\rm{FDFD}}^{\Delta x = \lambda/240}}|$ averaged over angles and ridge widths is 0.11 radian.
So, we use $\Delta x = \lambda/40$ for the metasurface simulations using APF and using MaxwellFDFD.

Similarly, Figure~\ref{fig:dis_error}b shows the discretization error of the transmission phase shift for RCWA, $\Phi_{\rm{RCWA}}^{{\rm numG}=5}  - {\Phi_{\rm{RCWA}}^{{\rm numG}=99}}$, where ${\rm numG}$ is the number of Fourier components used in the unit cell simulations.
The error is comparable to Fig.~\ref{fig:dis_error}a, with averaged $|\Phi_{\rm{RCWA}}^{{\rm numG}=5}  - {\Phi_{\rm{RCWA}}^{{\rm numG}=99}}|$ being 0.10 radian.
So, we use 5 Fourier components per unit cell (11 Fourier components per $\lambda$) for the metasurface benchmarks with RCWA.

We checked that FDFD and RCWA give consistent results, with $|{\Phi_{\rm{FDFD}}^{\Delta x = \lambda/240}}  - {\Phi_{\rm{RCWA}}^{{\rm numG}=99}}|$ averaging to
 0.03 radian.

\section{Captions for supplementary movies\label{sec:movies}}

\noindent {\bf Movie S1.}
Intensity profile of light transmitted through the mm-wide hyperbolic metalens as the incident angle varies. The profiles are normalized such that the incident flux is the same for all incident angles, and the colorbar is saturated near normal incidence in order to show the profiles at oblique incidence. The Strehl ratio and transmission efficiency is also shown.

\smallskip
\smallskip
\smallskip

\noindent {\bf Movie S2.}
Corresponding intensity profiles for the quadratic metalens.

%\clearpage

\bibliographystyle{naturemag}
\bibliography{supplementary}